\documentclass[%
 aip,
 amsmath,amssymb,
preprint,%
]{revtex4-1}

\usepackage{graphicx}
\usepackage{dcolumn}
\usepackage{bm}

\usepackage[utf8]{inputenc}
\usepackage[T1]{fontenc}
\usepackage{mathptmx}
\usepackage{etoolbox}
\usepackage{setspace}
\usepackage{subfigure}
\usepackage{CJKutf8}
\usepackage{color}

\makeatletter
\def\@email#1#2{%
 \endgroup
 \patchcmd{\titleblock@produce}
  {\frontmatter@RRAPformat}
  {\frontmatter@RRAPformat{\produce@RRAP{*#1\href{mailto:#2}{#2}}}\frontmatter@RRAPformat}
  {}{}
}%
\makeatother
\begin{document}
\begin{CJK*}{UTF8}{gbsn}

\preprint{AIP/123-QED}

\title{The stability analysis based on viscous theory of Faraday waves in Hele-Shaw cells}
\author{Xingsheng Li (李兴盛)}
\author{Jing Li (李靖)}
 \email{lijing\_@sjtu.edu.cn}
\affiliation{ 
Marine Numerical Experimental Center, State Key Laboratory of Ocean Engineering, School of Naval Architecture, Ocean and Civil Engineering, Shanghai Jiao Tong University, Shanghai, 200240, PR China}%

\date{\today}
             
\begin{abstract}
The linear instability of Faraday waves in Hele-Shaw cells is investigated with consideration of the viscosity of fluids after gap-averaging the governing equations due to the damping from two lateral walls and the dynamic behavior of contact angle. A new hydrodynamic model is thus derived and solved semi-analytically. The contribution of viscosity to critical acceleration amplitude is slight compared to other factors associated with dissipation, and the potential flow theory is sufficient to describe onset based on the present study, but the rotational component of velocity can change the timing of onset largely, which paradoxically comes from the viscosity. The model degenerates into a novel damped Mathieu equation if the viscosity is dropped with two damping terms referring to the gap-averaged damping and dissipation from dynamic contact angle, respectively. The former increases when the gap size decreases, and the latter grows as frequency rises. When it comes to the dispersion relation of Faraday waves, an unusual detuning emerges due to the imaginary part of the gap-averaged damping.
\end{abstract}

\maketitle
\end{CJK*}

\section{\label{sec:level1}INTRODUCTION}

When a container filled with liquids is subjected to vertical vibrations, Faraday waves are formed at the free surface in some specific conditions. This phenomenon has been widely applied to numerous areas of science and industry, such as assembly of microscale beads,\cite{chen2014microscale} atomization,\cite{liu2019experimental} inkjet printing,\cite{turkoz2019reduction} oil-gas separator on floating offshore platforms\cite{faltinsen2017sloshing} and so on. This type of resonant surface wave was first experimentally observed by Faraday\cite{faraday1831peculiar} in 1831 and subsequently by \citet{matthiessen1868akustische} and \citet{rayleigh1883vii}. From then on, a large number of researchers were attracted by this interesting physical process and devoted themselves to revealing the mechanism behind it via experiments, numerical simulations, and theoretical analyses. Among these studies, linear instability and pattern selection of Faraday waves have received the most attention. The linear instability corresponds to the critical conditions when Faraday waves occur, which is called the onset. Pattern selection concentrates on which form Faraday waves will appear in. Depending on the fluid properties and parameters of external excitation, different wave patterns have been observed in experiments, which include triangles,\cite{muller1993periodic} parallel stripes, squares, hexagons and higher symmetry quasi-patterns.\cite{edwards1994patterns} Spatio-temporal chaos\cite{kudrolli1996patterns} and highly localized circular waves\cite{lioubashevski1996dissipative} may also emerge. These complicated phenomena and nonlinearity bring challenges to mathematical modelings and numerical simulations.

To give the prediction of linear instability, \citet{benjamin1954stability} applied the theory of Mathieu equation. Results reflected that Faraday waves of half-frequency and synchronous response alternately occurred. But this theoretical method is only valid in inviscid restrictions. In the limit of low viscosity, \citet{landau1959fluid} introduced a very thin layer at the free surface, while the rest of the flow remained potential. There was no other signal progress until a full hydrodynamic system was developed by \citet{kumar1994parametric} to solve the stability problem for the interface between two fluids of arbitrary viscosity. The derivation began with the Navier-Stokes equations and the effect of viscosity was completely retained. Results showed a great agreement with experimental data of \citet{edwards1993parametrically}, which confirmed that the viscosity of fluids was essential in the stability analysis of Faraday waves. Since then, accurate prediction in the case of three-dimensional Faraday waves has been achieved. Inspired by them, \citet{pototsky2016faraday} studied the linear instability problem of a two-layer liquid film with a deformable upper surface.

Compared to the prediction of instability thresholds, pattern selection is more complicated because of the nonlinearity and resonant interactions of wave vectors. Taking the limitation of linear analysis into account, nonlinear approximations are necessary to describe the pattern formation beyond the instability threshold, among which the derivation of amplitude equations has become a classic theoretical method. Early analyses of pattern mechanism were based on the assumption that the bulk flow was irrotational and the Hamiltonian description was applied.\cite{miles1984nonlinear,miles1993faraday} The viscous dissipation of bulk flow was simply treated as small damping and represented explicitly by the rate of energy loss which had a form of integration.\cite{milner1991square} From the comparison with experiments of \citet{douady1988pattern}, it is obvious that the above assumption about dissipation is not appropriate and the contribution of the free surface to the damping should be considered. Focused on the dissipation near the surface skin, \citet{zhang1997pattern} used a quasi-potential approximation to describe the small vortical layer near the free surface. However, their theory is limited to fluid systems with small viscous dissipation. \citet{chen1999amplitude} then extended this previous research to arbitrary viscosity by following the earlier work of \citet{kumar1994parametric}. The difference was that nonlinear terms in the system of equations were retained. The theoretical results were validated by \citet{westra2003patterns} later, which further explained the importance of viscosity in Faraday waves. Recently, \citet{zhang2023pattern} applied modal decomposition to reveal the nonlinear mechanisms behind pattern formation in a brimful cylinder. This further highlighted the nonlinear feature of Faraday waves. In this sense, numerical simulations give more information about the nonlinear viscous problem of Faraday waves. However, most simulations have been two-dimensional, such as \citet{murakami2001two} and \citet{ubal2003numerical}. Until 2009, \citet{perinet2009numerical} presented a systemic simulation about the full dynamics of Faraday waves in a three-dimensional field for two viscous fluids. The critical accelerations and wavenumbers, different patterns, and velocity fields were reproduced.

In Hele-Shaw cells, however, the study of Faraday waves has greatly changed. On the one hand, more complicated phenomena have been observed in experiments. In 2011, \citet{rajchenbach2011new} found a new type of solitary wave in a Hele-Shaw cell which was highly localized and oscillated periodically. \citet{li2014observations} then conducted a series of more extensive experiments and discovered a new wave structure with multiple crests and troughs. Later, a family of two-dimensional Faraday waves with plump crests and flat troughs almost getting in touch with the bottom of the container was observed in extremely shallow depth.\cite{li2015observation}  And two coupled Faraday waves were observed in a covered Hele-Shaw cell filled with two immiscible liquids with a free upper surface.\cite{li2018observation} The upper one vibrates vertically, while the crests of the lower one oscillate horizontally with unchanged wave heights. These novel Faraday waves show a strong nonlinear property and some may only appear in very localized areas, which makes the physical phenomena more complex to model. On the other hand, the dissipation is strengthened in Hele-Shaw flows. In a traditional Faraday resonance system, it mainly arises from the viscosity of fluids. \citet{milner1991square} found that dissipation at the walls of the container was negligible compared to the bulk contribution and there was no contribution to dissipation from a moving contact line. But for Hele-Shaw cells with extremely narrow gaps, the no-slip condition is often introduced and the dissipation from two lateral walls is enhanced largely. Meanwhile, the meniscus formed along the gap direction has an apparent contribution to the curvature because of the diminutive gap size, which results in larger surface contributions to dissipation. These two dissipation sources increase the instability threshold significantly. Hence, the prediction of Faraday instability becomes more complex to a large extent in Hele-Shaw cells and has been the focus of research.

To obtain the thresholds of Faraday instability in Hele-Shaw cells, \citet{rajchenbach2011new} introduced the so-called \textit{Rayleigh’s external viscosity} $\gamma=12\nu/b^2$ to calculate the dissipation resulting from the narrow gap, with $\nu$ the fluid kinematic viscosity and $b$ the cell’s gap-size. However, the surface tension is absolutely neglected, which is essential for such gravity-capillary waves. \citet{li2018faraday} assumed that the flow through the gap was of Poiseuille type and developed the so-called gap-averaged Navier-Stokes equations which included the viscous term combining with gravity and capillary force. An open-source code Gerris\cite{popinet2003gerris,popinet2009accurate} was then used to solve the above equations numerically. The wave profile, wave height together with flow field information were obtained and compared with experimental data. A scaling law with respect to the Strouhal number was proposed for the first time to predict the wave height. Then, \citet{li2019stability} established a gap-averaged model to include surface tension and meniscus between the two side walls. Nonlinear terms in equations were reserved and the Lyapunov's first method was used to obtain the threshold of onset. From the comparison with experiments, it is obvious that the gap-averaged model gives a better prediction of Faraday instability than previous work. \citet{martino2020sediment} discussed the influence of liquid depth on Faraday instability and a new scaling law of wave height was given later.\cite{10.1063/5.0128809} But it only works when critical acceleration amplitude is known as a priori information, which cannot be calculated accurately without laboratory measurements. Besides, only gravity is involved in this scaling law, yet the capillary effect has a similar weight on the Faraday instability problem.\cite{Li2023response} The influence of the free surface was further verified by \citet{li2022numerical}. Focusing on the meniscus formed at the free surface, the impact of contact line with hysteresis on the Faraday instability was studied. The energy dissipation with respect to viscosity and surface tension was analyzed for three different boundary conditions at the lateral walls. In conclusion, they emphasized that the contact line motion should not be ignored because it largely delayed the onset of Faraday waves. Inspired by the theoretical work of \citet{li2019stability}, \citet{rachik2023effects} studied the effects of finite depth and surface tension on the dispersion relation and Faraday instability. More recently, \citet{bongarzone2023revised} modified the damping coefficient resulting from the gap-averaging process with the theory of oscillating Stokes boundary layer. The results indicate that this theory may fit Hele-Shaw flows better than the Darcy approximation used by \citet{li2019stability}.

Although many works concentrate on the linear instability of Faraday waves in Hele-Shaw cells, the above mathematical analyses are all based on potential assumptions or have neglected viscous terms in equations. The effect of viscous dissipation from the bulk flow to the Hele-Shaw system is still unclear. Numerical simulations are also limited since the instability threshold cannot be provided directly. A prediction of Faraday instability in Hele-Shaw cells based on viscous theory is necessary to reveal the influence of viscosity, which is the motivation of the present work. In this paper, the stability of Faraday waves in Hele-Shaw cells is studied theoretically with the viscosity of fluids being reserved. The contribution of viscosity to critical conditions of onset is revealed. 
In accordance with \citet{bongarzone2023revised}, the gap-averaging strategy is updated. 
In addition, a novel damped Mathieu equation is derived in the absence of viscosity. 
Before elaborating, the dissipation in such a Hele-Shaw system should be defined. The onset of Faraday waves has been thought to depend on the dissipation of the entire system. When the balance between the external driving force and dissipation is broken, the free surface loses its stability and Faraday waves appear. 
The dissipation of liquid motions in Hele-Shaw cells falls into four categories: the two-dimensional viscous dissipation referring to the effect of viscosity, the gap-averaged damping resulting from two lateral walls, the dissipation from dynamic contact angle which describes the interaction of the free surface and lateral walls, and dissipation from two ends of the rectangular container.
Among them, the first one is normally overlooked by previous studies,\cite{li2019stability,rachik2023effects,bongarzone2023revised} which are limited to inviscid fluids as has been mentioned at the beginning of this paragraph. 
With regard to the dissipation arising from two ends, from experiments conducted in a thin annular container by \citet{bongarzone2023revised}, the influence of two ends of the vessel is not a main reason for distinctions between theory and experiment, so we ignore it in the present paper.

The remainder of this paper is organized as follows. In Sec.~\ref{sec:level2}, a detailed derivation of the mathematical model with viscosity and capillary effect being included is described. Then it is solved semi-analytically based on linear stability theory in Sec.~\ref{sec:level3}. To highlight the effect of viscosity, a damped Mathieu equation is obtained based on potential flow theory in Sec.~\ref{sec:level4}, which only contains two main sources of dissipation resulting from the gap-averaged damping and dynamic contact angle. In Sec.~\ref{sec:level5}, The findings, along with a comparison to experimental data, are presented. Additionally, the examination of the effects of viscosity and the remaining dissipation is discussed. Finally, some concluding remarks are given in Sec.~\ref{sec:level6}.
\begin{figure}
\includegraphics[width=0.8\linewidth]{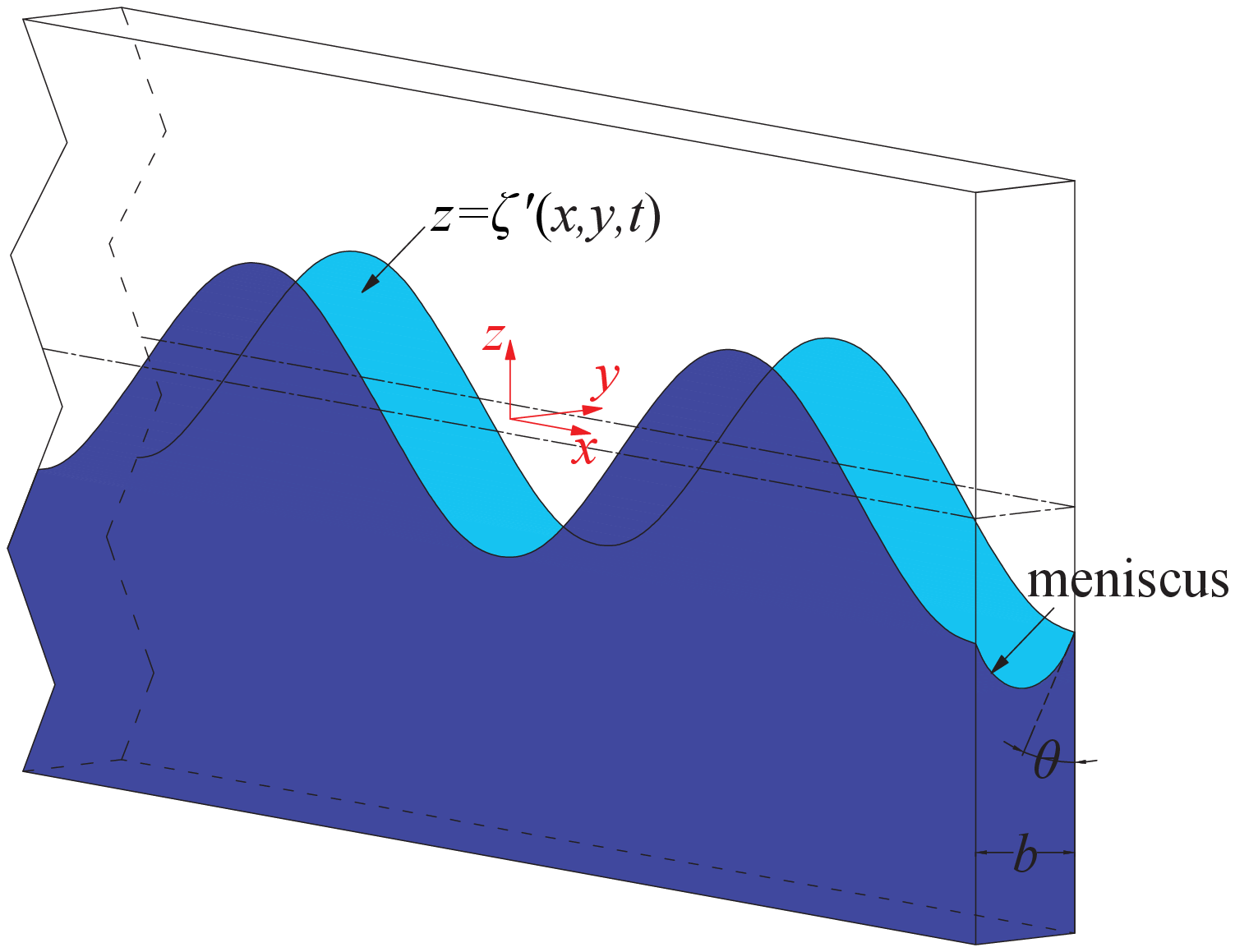}
\caption{\label{fig:1} Sketch of Faraday waves in a Hele-Shaw cell. Here \textit{b} denotes the gap size of the Hele-Shaw cell and $\theta$ is the contact angle of the liquid on the lateral wall.}
\end{figure}

\section{\label{sec:level2}MATHEMATICAL MODEL}

We consider here a horizontally infinite Hele-Shaw cell which undergoes a vertical periodic oscillation. The vessel is of width \textit{b} and filled with a kind of liquid of density $\rho$ and dynamic viscosity $\mu$. In the Cartesian coordinate system as shown in Fig.~\ref{fig:1} which moves with the oscillating vessel, $z=0$ is set at the flat free surface when at rest. The oscillation is sinusoidal with acceleration amplitude \textit{a} and angular frequency $\omega=\frac{2\pi}{T}$ (\textit{T}: the forcing period). What we are concerned about is the threshold of acceleration corresponding to the onset of Faraday waves with frequency fixed. Since the free surface is flat overall before Faraday waves occur, a linear theory is sufficient to obtain the critical conditions. This has been verified by \citet{kumar1994parametric} and \citet{chen1999amplitude} in three-dimensional Faraday instability problem and by \citet{bongarzone2023revised} in the Hele-Shaw scenario, where linear methods such as the Floquet theory were used.

Although the gap size \textit{b} is sufficiently small compared to the wavelength $\lambda$,\cite{li2019effect} which means the flow in Hele-Shaw cells can be regarded as two-dimensional, the dissipation resulting from two lateral walls cannot be ignored.
Hence, a reasonable gap-averaging process is necessary.

\subsection{\label{sec:level2-1}Governing equation}

To derive a gap-averaged model, there are two strategies in previous works. 
One is to gap-average the Navier-Stokes equations first, and then the flow is assumed to be inviscid and irrotational, and hence the potential flow theory is applied. 
As a consequence, a damping coefficient appears naturally in the boundary condition at the free surface.\cite{li2019stability} 
The other one is that the velocity rather than the equations is the objective to be gap-averaged.
After plugging the Floquet ansatz into the normal stress boundary condition, the damping terms once again emerge at the free surface.
However, this strategy is valid only if viscous terms in Navier-Stokes equations are neglected.\cite{bongarzone2023revised} 
Here, one of the major motivations is to examine the outcome with the viscous effect involved, so a different strategy has been proposed in the present study. 
Although there is no reference concentrating on viscosity in the case of Hele-Shaw flows, viscous theory has been successfully developed in three-dimensional Faraday waves.\cite{kumar1994parametric,chen1999amplitude}
Based on it, the governing equation is simplified to two-dimensional with no assumption on viscosity.

Under vertical periodic vibration, the equations governing fluid motion are
\begin{gather}
\partial _t\bm{u}'+\left (\bm{u}'\cdot \nabla  \right ) \bm{u}' =-\frac{1}{\rho } \nabla p+\nu \nabla^2\bm{u}'-g_z\left ( t \right ) \bm{e}_z
\label{eq:1},
\\
\nabla \cdot \bm{u}'=0
\label{eq:2},
\end{gather}
where $\bm{u}'=\left ( u',v',w' \right ) $ is the fluid velocity in \textit{x}, \textit{y}, \textit{z}-directions, \textit{p} is the pressure, $\nu=\frac{\mu}{\rho}$ is kinetic viscosity, $g_z\left ( t \right )=g-a\cos\omega t$ is the effective gravity under sinusoidal oscillation with \textit{g} denoting gravity. The state of rest is a quiescent fluid with a pressure distribution $p=-\rho g_z\left ( t \right )z$. By applying $-\left ( \nabla\times \nabla\times \right )$ to Eq.~(\ref{eq:1}), where $\nabla\times$ denotes taking the curl, together with the continuity condition expressed by Eq.~(\ref{eq:2}), the pressure term can be eliminated. Then the velocity satisfies
\begin{equation}
\partial _t\nabla^2\bm{u}'-\nu\nabla^2\nabla^2\bm{u}'=\nabla\times \nabla\times\left ( \bm{u}'\cdot \nabla \right )\bm{u}'
\label{eq:3}.
\end{equation}
Linearizing the above equation and taking the component in the \textit{z}-direction, one can obtain
\begin{equation}
\left ( \partial _t-\nu\nabla^2 \right ) \nabla^2w'=0,
\label{eq:4}
\end{equation}
where $\nabla^2=\partial _{xx}+\partial _{yy}+\partial _{zz}$ is the Laplacian.

Eq.~(\ref{eq:4}) is still three-dimensional. Simplifying this equation to two-dimensional is necessary. Due to the consideration of viscosity, we have taken higher-order derivatives of Navier-Stokes equations to eliminate the pressure term, which refers to the derivation from Eq.~(\ref{eq:1}) to Eq.~(\ref{eq:3}). 
This results in several terms associated with $\partial_{xx}$ or $\partial_{yy}$ in Eq.~(\ref{eq:4}). 
To deal with these terms, in three-dimensional circumstances, a horizontally infinite plane was considered,\cite{kumar1994parametric} whose normal modes were trigonometric functions, e.g. $\sin \left (\bm{k} \cdot\bm{x} \right ) $, with $\bm{k}=\left(k_x,k_y\right)$ and $\bm{x}$ denoting horizontal wave vector and \textit{x},\textit{y}-components, respectively. According to this assumption, there exists a mathematical relationship that $\partial _{xx}=-k_x^2$ and $\partial _{yy}=-k_y^2$.
However, in Hele-Shaw cells, there is no wave profile generated in the direction of the gap except the meniscus. 
So $k_y=0$ and $\nabla^2$ is replaced by $\partial _{xx}+\partial_{zz}$ in Eq.~(\ref{eq:4}). 
If the velocity profile in the \textit{y}-direction is introduced into  Eq.~(\ref{eq:4}) and followed by the gap-averaging process, the gap-averaged damping is generated from terms associated with $\partial _{yy}$, which is contradicted with the calculation of wavenumber and will lead to mistakes in the dispersion relation. 
In fact, the governing equation of previous theoretical studies has been a two-dimensional Laplace equation, where no gap-averaged damping is included.
So it is reasonable to remove the term $\partial_{yy}$ in the Laplacian $\nabla^2$ during the whole theoretical process to guarantee a right dispersion relation, except the derivation of normal stress boundary condition, where the gap-averaged damping is introduced. 
The final governing equation is
\begin{equation}
\left [ \partial _t-\nu \left ( \partial _{xx}+\partial _{zz}\right ) \right ] \left ( \partial _{xx}+\partial _{zz}\right )w=0.
\label{eq:5}
\end{equation}
In Eq.~(\ref{eq:5}) and what follows, \textit{w} is a function of \textit{x}, \textit{z}, \textit{t}. The term $\sin \left ( k x \right )$ can be separated from \textit{w}, with $k$ denoting the wavenumber in the \textit{x}-direction, where we have omitted the subscript \textit{x} for brevity.

\subsection{\label{sec:level2-2}Boundary conditions}

To obtain boundary conditions for Faraday instability in Hele-Shaw cells, the effect of two lateral walls should be defined first, which is the main source of damping in the Hele-Shaw system. Now that the equations will be reduced to two-dimensional, a typical method is to introduce a velocity profile in the \textit{y}-direction. To deal with it, Li \textit{et al.} applied the Kelvin--Helmholtz--Darcy theory\cite{gondret1997shear} and assumed that the flow through the gap was of Poiseuille type.\cite{li2019stability,li2018faraday} Based on this, the velocity profile was supposed to remain a parabolic shape in the \textit{y}-direction. The three-dimensional Navier-Stokes equations were then integrated across the gap and averaged with gap size \textit{b}. 
However, the oscillation of the vessel is not considered, which may result in the unsteady nature of such Hele-Shaw flows. When studying the sloshing in Hele-Shaw cells, \citet{viola2017sloshing} included the oscillating Stokes boundary layer which focused on the unsteady terms of Navier-Stokes equations. 
\citet{bongarzone2023revised} then extended this approach to the Faraday instability problem and modified it to fit Floquet theory. 
In order to match the present study, the velocity profile can be derived as
\begin{equation}
w'\left ( x,y,z,t \right ) =\left \{ 1-\frac{\cosh \left [ \left ( 1+\mathrm{i} \right ) y/(b\delta _j) \right ] }{\cosh \left [ \left ( 1+\mathrm{i} \right )/2\delta _j \right ] }  \right \} w\left ( x,z,t \right ),
\label{eq:6}
\end{equation}
where i is the imaginary unit, $j=1,3,5,\cdots$ corresponds to each Fourier component in the ansatz (Eq.~(\ref{eq:24}) in the next section), and $\delta _j=\frac{\sqrt{2\nu /\omega } }{b\sqrt{j/2}}$ denotes the thickness of the oscillating Stokes boundary layer after being nondimentionalized by \textit{b}. A detailed derivation of velocity profile Eq.~(\ref{eq:6}) is shown in Appendix~\ref{sec:levelA1}. We emphasize that the velocity distribution through the thin gap follows the expression of Eq.~(\ref{eq:6}).

To obtain boundary conditions satisfied at the moving free surface, the mathematical description of the surface has to be given first. As shown in Fig.~\ref{fig:1}, the real wave profile contains two parts:\cite{li2019stability}
\begin{equation}
\zeta' \left ( x ,y,t\right ) =\zeta \left ( x,t \right ) +\frac{y^2}{b} \cos \theta,
\label{eq:7}
\end{equation}
where the two terms on the right-hand side indicate the principal wave profile observed in the \textit{x}-direction and the meniscus formed between two lateral walls, respectively. The latter is approximated by assuming that the meniscus is a segment of circular and the parabolic function is obtained based on Taylor expansion. Then according to \citet{chen1999amplitude}, the outward-pointing normal unit vector $\bm{n}$ is written as
\begin{equation}
\bm{n}=\frac{\left ( -\partial _x\zeta , -\partial _y\zeta,1\right ) }{\left [ 1+ \left ( \partial _x\zeta \right )^2 +\left ( \partial _y\zeta \right ) ^2\right ]^{1/2} }.
\label{eq:8}
\end{equation}
And two linearly independent tangential unit vectors $\bm{t}_1$ and $\bm{t}_2$ are
\begin{equation}
\bm{t}_1=\frac{\left ( 1 , 0,\partial _x\zeta\right ) }{\left [ 1+ \left ( \partial _x\zeta \right )^2  \right ]^{1/2} }, \qquad 
\bm{t}_2=\frac{\left ( 0 , 1,\partial _y\zeta\right ) }{\left [ 1+ \left ( \partial _y\zeta \right )^2  \right ]^{1/2} }.\label{eq:9}  
\end{equation}

The kinematic boundary condition at the free surface is
\begin{equation}
\left [ \partial _t+\left ( \bm{u}\cdot \nabla _H \right )  \right ] \zeta '=w', \quad \text{at } z=\zeta,\label{eq:10}
\end{equation}
with $\nabla _H=\bm{e}_x\partial _x+\bm{e}_y\partial _y$. Since the meniscus is too small compared to the principal wave profile, the position of the free surface is set to $z=\zeta$ and only $\zeta\left(x,t\right)$ is considered in $\zeta'$ on the left-hand side. Eq.~(\ref{eq:10}) is then Taylor expanded around $z=0$. After eliminating the nonlinear terms, integrating along \textit{y}, and computing its mean divided by \textit{b}, it has a form of
\begin{equation}
\partial _t\zeta =\lambda_j w, \quad \text{at } z=0.\label{eq:11}
\end{equation}
where the coefficient $\lambda_j= 1+\left ( i-1 \right ) \delta_j \tanh \left [ \left ( 1+i \right ) /\left (2\delta_j\right )  \right ]$ is generated from the gap-averaging process with the expression of $w'$ shown in Eq.~(\ref{eq:6}).

The remaining boundary conditions at the free surface are derived by considering the stress tensor, whose components are $T_{ij}=-\left [ p-\rho g_z\left ( t \right ) z \right ] \delta _{ij}+\mu \left ( \partial _ju'_i+\partial _iu'_j \right)$ with \textit{i}, \textit{j} referring to \textit{x}, \textit{y}, \textit{z}. Now that the effect of the air phase above the liquid is neglected, the tangential components of the stress tensor are zero at the free surface, which means
\begin{equation}
\bm{t}_m\cdot \bm{T} \cdot \bm{n}=0, \quad \text{at } z=\zeta,\label{eq:12}
\end{equation}
with $m=1,2$. There are actually two equations in Eq.~(\ref{eq:12}), but by taking the \textit{x}-derivatives and \textit{y}-derivatives respectively, along with the continuity condition of Eq.~(\ref{eq:2}), a combination is obtained. After being expanded around $z=0$ and linearized, Eq.~(\ref{eq:12}) reads
\begin{equation}
\left ( \nabla_H^2-\partial_{zz} \right )w'=0, \quad \text{at } z=0.\label{eq:13}
\end{equation}
By using the same method as that implemented to derive Eq.~(\ref{eq:5}) when dealing with $\nabla_H^2$, the tangential stress boundary condition Eq.~(\ref{eq:13}) is simplified to
\begin{equation}
\left (\partial_{xx}-\partial_{zz} \right )w=0, \quad \text{at } z=0.\label{eq:14}
\end{equation}

The normal stress at the free surface is balanced by capillarity force, which means
\begin{equation}
\bm{n} \cdot \bm{T} \cdot \bm{n}=\sigma \kappa, \quad \text{at } z=\zeta,\label{eq:15}
\end{equation}
where $\sigma$ is surface tension coefficient and $\kappa$ is surface curvature. 
With the free surface elevation of Eq.~(\ref{eq:7}), the curvature is then expressed by
\begin{equation}
\kappa =\frac{\partial _{xx}\zeta }{\sqrt{1+\left ( \partial_x\zeta  \right ) ^2} } +\frac{2}{b} \cos \theta.\label{eq:16}
\end{equation}

Before we move on to further derivation, the term relevant to the contact angle in Eq.~(\ref{eq:16}) should be clarified first. 
As has been mentioned by \citet{li2019stability}, most of the studies have assumed that the out-of-plane interface shape is semicircular and naturally $\theta=180^{\circ}$. However, in the presence of vertical periodic vibration applied to the Hele-Shaw cell, the surface near the lateral walls undergoes an up-and-down process. This means the contact angle changes constantly and a dynamic contact angle model should be considered. Following \citet{li2019stability}, a linear model is introduced and the cosine of $\theta$ is evaluated as\cite{hamraoui2000can}
\begin{equation}
\cos \theta=1-\frac{\beta}{\mu}Ca,\label{eq:17}
\end{equation}
where $\beta$ is friction coefficient and $Ca=\frac{\mu}{\sigma}w'$ is the capillary number defined using vertical velocity at the free surface $w'=\partial_t \zeta$. This model can be interpreted as that the static contact angle is $0^{\circ}$ and a modification associated with velocity is introduced with the consideration of the dynamic behavior of the free surface. Only the term that varies with \textit{Ca} works after the linearizing process. The coefficient $\beta$ is a phenomenological parameter which has been measured experimentally by \citet{hamraoui2000can} and successfully used in previous studies.\cite{li2019stability,bongarzone2023revised}

Substituting Eq.~(\ref{eq:16}) and Eq.~(\ref{eq:17}) into the normal stress boundary condition Eq.~(\ref{eq:15}), a linearized equation is obtained after removing nonlinear terms:
\begin{equation}
-p+2\mu \partial _zw'+\rho \left ( g-a\cos \omega t \right ) \zeta -\sigma \partial_{xx}\zeta +\frac{2\beta}{b}\partial_t \zeta=0, \quad \text{at } z=0. \label{eq:18}
\end{equation}
Then Eq.~(\ref{eq:18}) is gap-averaged with the consideration of Eq.~(\ref{eq:6}), which leads to
\begin{equation}
 -p+2\mu \lambda_j \partial _zw +\rho \left ( g-a\cos \omega t \right ) \zeta -\sigma \partial_{xx}\zeta +\frac{2\beta}{b}\partial_t \zeta=0, \quad \text{at } z=0. \label{eq:19}
\end{equation}
To eliminate the pressure term, we first take the horizontal divergence of Eq.~(\ref{eq:1}). With the consideration of continuity condition Eq.~(\ref{eq:2}), a linearized equation is obtained that
\begin{equation}
-\partial _{zt}w'+\nu \nabla^2\partial_zw'=-\frac{1}{\rho}\nabla_H^2p. \label{eq:20}
\end{equation}
Also Eq.~(\ref{eq:20}) is gap-averaged with velocity profile Eq.~(\ref{eq:6}) being used
\begin{equation}
-\frac{2\left ( 1+\mathrm{i} \right )\nu \tanh \left [ \left ( 1+\mathrm{i} \right ) /2\delta_j  \right ] }{b^2\delta_j }\partial _zw- \lambda_j\bigr [ \partial _{zt}w- \nu \bigl ( \partial _{xxz}w+\partial _{zzz}w \bigl )  \bigr ]=-\frac{1}{\rho}\partial_{xx}p. \label{eq:21}
\end{equation}
By operating Eq.~(\ref{eq:19}) with $\partial_{xx}$ and combining with Eq.~(\ref{eq:21}), the pressure term is eliminated
\begin{widetext}
\begin{equation}
\begin{split}
-\frac{2\left ( 1+\mathrm{i} \right )\nu \tanh \left [ \left ( 1+\mathrm{i} \right ) /2\delta_j  \right ] }{b^2\delta_j }\partial _zw-  \lambda_j\left [ \partial _{zt}w-\nu \left ( 3\partial _{xxz}w+\partial _{zzz}w \right )  \right ]  +\left ( g-a\cos \omega t \right ) \partial _{xx}\zeta \\-\frac{\sigma }{\rho } \partial _{xxxx}\zeta +\frac{2\beta}{\rho b} \partial _{xxt} \zeta =0, \quad \text{at } z=0.\label{eq:22}
\end{split}
\end{equation}
\end{widetext}

Before we move on to the next part, some points need to be emphasized. 
Our attentive readers will notice that there is an alternative way to obtain Eq.~(\ref{eq:22}) through having the three-dimensional normal stress boundary condition first and then gap-averaging it.
However, this treatment will lead to misunderstanding and confusion about the concepts of the two terms in the present context: viscous dissipation and gap-averaged damping.
As mentioned in Sec.~\ref{sec:level1}, the former one indicates the in-plane viscosity and the latter one stems from the out-of-plane counterpart.
In Hele-Shaw cells, the latter one occupies a significant proportion in dissipation and has been successfully calculated in previous works based on potential flow theory,\cite{li2019stability,rachik2023effects} where a damping coefficient is obtained arising from the \textit{y}-component of viscous terms in Eq.~(\ref{eq:1}) through the gap-averaging process. 
Please note that viscosity is still the very source of the gap-averaged damping. 
So far the former one does not appear due to the use of potential flow theory for the in-plane system.
Nevertheless, if we follow the derivation in three-dimensional Faraday instability problem in the first place,\cite{kumar1994parametric,chen1999amplitude} where viscosity in each dimension exists, the linearized form of Eq.~(\ref{eq:15}) will be operated with $\nabla_H^2=\partial_{xx}+\partial_{yy}$ and Eq~(\ref{eq:1}) will be operated with $(\nabla_H\cdot)$ to eliminate pressure. 
A term $3\nu \nabla_H^2\partial_zw'$ will be obtained as a part of viscous dissipation after that and ends up the gap-averaged damping after the gap-averaging process. 
It will genuinely overestimate the damping since this procedure mixes up different ingredients of viscosity together and abuses the gap-averaging to all of them.
The truth is that viscosity in Navier-Stokes equations is associated with the bulk flow which should be simplified for the $y$-component, while the counterpart in the original normal stress boundary condition is relevant to the dissipation from the free surface which should be treated as the in-plane factor rather than gap-averaged one. 
To avoid this interference in the calculation of damping, we have gap-averaged Eq.~(\ref{eq:18}) and Eq.~(\ref{eq:20}) individually first and then eliminate the pressure term. 
In this sense, the out-of-plane viscosity transforms into the gap-averaged damping, and the viscous dissipation is deemed as in-plane in Hele-Shaw cells.
Eq.~(\ref{eq:22}) is thus the final form.

For the bottom wall, \citet{martino2020sediment} studied the influence of liquid depth $\Delta h$ on the threshold of Faraday instability and concluded that for $\frac{\Delta h}{b}\gtrapprox2$ the threshold tends to asymptotic values. In experiments conducted in Hele-Shaw cells,\cite{li2019stability} this ratio far exceeded the critical valve. In this sense, assuming that the liquid layer extends to $z=-\infty$ is feasible. A null condition at $z=-\infty$ means
\begin{equation}
w=0, \quad \text{at } z=-\infty.\label{eq:23}
\end{equation}

The above equations of (\ref{eq:5}),(\ref{eq:11}),(\ref{eq:14}),(\ref{eq:22}), and (\ref{eq:23}) constitute the final hydrodynamic equations for Faraday instability in Hele-Shaw cells, in which the solutions to be determined are vertical velocity $w\left ( x,z,t \right )$ and surface displacement $\zeta\left ( x,t \right )$.

\section{\label{sec:level3}SOLUTION BASED ON LINEAR STABILITY THEORY}

As has been mentioned, a linear solution is sufficient to describe the instability of the free surface.
This has been validated in the three-dimensional scenario by \citet{kumar1994parametric}. 
Their formulation of deriving a system of equations to define the instability threshold was briefly revisited by \citet{chen1999amplitude}. The difference was that the generalized eigenvalue problem was rewritten as an equivalent, implicit expression for the acceleration threshold. When wavenumber \textit{k} is determined, the threshold value $a_c$ can be solved by algebraic iteration. 
A similar procedure is extended to Hele-Shaw cells as follows.

Faraday waves in Hele-Shaw cells possess a subharmonic motion at a frequency of $\omega/2$ observed in previous experiments.\cite{li2018observation} 
Only subharmonic cases are therefore considered here, which is reflected in the exponential part of the solutions. 
Besides, if the fundamental mode $e^{\mathrm{i} \omega t/2}$ is included alone, the stability analysis will be only appropriate for small viscosity, which has been illustrated in Ref.~\onlinecite{chen1999amplitude}. 
With the consideration of these, the Fourier series is applied to contain all the harmonics of the fundamental mode $e^{\mathrm{i} \omega t/2}$. 
The ansatz for $w\left ( x,z,t \right )$ and $\zeta\left ( x,t \right )$ are then written as
\begin{equation}
\begin{split}
w\left ( x,z,t \right ) &=\sin \left ( kx \right ) \sum_{j=1,3,5,\cdots } e^{\mathrm{i} j \omega t/2}w_j\left ( z \right )A_j+c.c.,\\
\zeta \left ( x,t \right ) &=\sin \left ( kx \right ) \sum_{j=1,3,5,\cdots } e^{\mathrm{i} j \omega t/2}A_j+c.c.,\label{eq:24}
\end{split}
\end{equation}
where $A_j$ are complex amplitudes, \textit{c.c.} denotes the complex conjugate.

Substituting the assumed solutions given by Eq.~(\ref{eq:24}) into the governing equation (\ref{eq:5}), one obtains an ordinary differential equation for $w_j\left ( z \right )$ as
\begin{equation}
\left [ \frac{1}{2} \mathrm{i} j\omega \left ( -k^2+\partial _{zz} \right ) -\nu \left ( -k^2+\partial _{zz} \right )^2  \right ]w_j\left ( z \right )=0.\label{eq:25}
\end{equation}
Together with the null condition (\ref{eq:23}), the solution of Eq.~(\ref{eq:25}) has a form of $w_j\left ( z \right )=m_je^{p_jz}+n_je^{q_jz}$ with $p_j=k$ and $q_j=\sqrt{k^2+\mathrm{i} j \omega /2\nu }$.

The kinematic boundary condition (\ref{eq:11}) and tangential stress boundary condition (\ref{eq:14}) are used to determine the coefficients $m_j$ and $n_j$, which are listed as
\begin{equation}
 m_j=\frac{2k^2\nu +\frac{1}{2} \mathrm{i} j\omega}{\lambda_j}, \qquad n_j=-\frac{2k^2\nu}{\lambda_j}.\label{eq:26}
\end{equation}

By now, the solutions expressed in Eq.~(\ref{eq:24}) are determined completely. The function $w_j\left ( z \right )$ is given by
\begin{equation}
w_j\left ( z \right )=\frac{1}{\lambda_j} \left [\left ( 2k^2\nu +\frac{1}{2} \mathrm{i} j\omega \right ) e^{kz}-2k^2\nu e^{z\sqrt{k^2+\mathrm{i} j \omega t/2\nu }}\right ]. \label{eq:27}
\end{equation}
According to Ref.~\onlinecite{chen1999amplitude} and private communication with Professor Jorge Vi\~{n}als ("One component of the velocity derives from a scalar potential, the other does not"), the two components on the right-hand side are irrotational and rotational components of the flow, respectively. Besides, the viscous and inviscid contributions are separated explicitly in the first component. 
It is easy to know that when it comes to the viscous effect, terms containing $\nu$ in Eq.~(\ref{eq:27}) should be all included.
There are two of them.
The first term of the first component can be considered as the irrotational part of the viscous dissipation, whereas the other one is a rotational correction to it. 
A detailed discussion of these two terms will be given in Sec.~\ref{sec:level5}.

The normal stress boundary condition serves as a solvability condition for the instability threshold $a_c$. 
Substituting the solutions into Eq.~(\ref{eq:22}), for each harmonic $e^{\mathrm{i} j\omega t/2}$, the equation should be satisfied.
Considering the equivalent relationship $a\cos \omega t=\frac{1}{2}a\left ( e^{\mathrm{i}\omega t} + e^{-\mathrm{i}\omega t}  \right )$ and the reality condition $A_{-1}=A_1^*$, equations for each Fourier component finally constitute a system of equations like
\begin{equation}
\begin{split}
&j=1:\quad H_1A_1-aA_1^*-aA_3=0,
\\&j=3:\quad H_3A_3-aA_1-aA_5=0,
\\&j=5:\quad H_5A_5-aA_3-aA_7=0,
\\&j=7:\quad \cdots,\label{eq:28}
\end{split}
\end{equation}
where $H_j$ are coefficients resulting from Eq.~(\ref{eq:22}), which is expressed explicitly as
\begin{widetext}
\begin{equation}
\begin{split}
H_j= \Big \{ \left [ \left ( 6k^2\nu +\mathrm{i} j \omega \right )\left ( m_jp_j+n_jq_j \right )  -2\nu \left ( m_jp_j^3+n_jq_j^3 \right )  \right ] \lambda_j+ 2k^2g + \frac{2k^4\sigma }{\rho} +\frac{2\mathrm{i}j\omega\beta k^2}{\rho b} \\ +\frac{4\nu \left ( 1+\mathrm{i}  \right )\left ( m_jp_j+n_jq_j \right ) \tanh \left [ \left ( 1+\mathrm{i} \right )/\left (2\delta _j \right ) \right ]  }{b^2 \delta _j} \Big \} /k^2.\label{eq:29}
\end{split}
\end{equation}
\end{widetext}

Eq.~(\ref{eq:28}) is a system of equations in unknown amplitudes $A_j$ of the solutions. 
When the working fluid properties, gap size \textit{b}, and vibrating angular frequency $\omega$ are all defined, $H_j$ becomes a function of wavenumber \textit{k} and acceleration amplitude \textit{a}. 
To obtain $a_c$ algebraically, we first truncate at finite terms \textit{n}. 
Then from the last equation, each $A_j$ in Eq.~(\ref{eq:28}) can be recursively formulated,
\begin{equation}
A_n=\frac{aA_{n-2}}{H_n}, \quad
A_{n-2}=\frac{aA_{n-4}}{H_{n-2}-a^2/H_n}, \quad
\dots ,\quad
A_3=\frac{aA_1}{H_{3}-a^2/H_5}.\label{eq:30}
\end{equation}
Until the first equation in Eq.~(\ref{eq:28}), it has the form of
\begin{equation}
\left ( H_1-\frac{a^2}{H_3-\frac{a^2}{H_5-\cdots } }  \right )A_1 -aA_1^*\equiv \bar{H_1}\left ( k,a \right ) A_1-aA_1^*=0.\label{eq:31}
\end{equation}
For convenience, we use $\bar{H}_1$ to represent the coefficient in front of $A_1$. For a given wavenumber \textit{k}, the instability threshold $a_k$ can be obtained implicitly from
\begin{equation}
a_k= \left | \bar{H_1}\left ( k,a_k \right ) \right |  =\left | H_1-\frac{a_k^2}{H_3-\cdots } \right |.\label{eq:32} 
\end{equation}

In practice, we have verified through convergence analysis that $n=11$ is sufficient to guarantee the accuracy of results. The value of wavenumber \textit{k} is first determined by traversing the values within the range from 0 to 1000 $(\rm{m^{-1}})$, which contains all experimental data in Hele-Shaw cells.\cite{kumar1994parametric} 
Then Eq.~(\ref{eq:32}) is iteratively solved, after which a critical acceleration amplitude is obtained as $a_k$. The smallest value of $a_k$ is the final instability threshold $a_c$, and the corresponding wavenumber is the critical wavenumber $k_c$.

\section{\label{sec:level4}POTENTIAL FLOW THEORY}

The above analysis has reserved the viscous effect of fluids. 
In this section, the potential flow theory is applied to clarify the impact of viscosity on Faraday instability in such Hele-Shaw cells. 
The resultant damped Mathieu equation actually can be obtained directly from the previous theory as long as all $\nu$ related terms are dropped in Eq.~(\ref{eq:27}).
For the sake of simplicity, the same equation is derived from the potential flow perspective.
It is interesting that if we look closely at the previous derivation, the inviscid but rotational flow cannot be kept intentionally and independently, which results in no difference between the ideal and potential flow perspectives.

Following \citet{li2019stability}, the linearized form of Eqs.~(\ref{eq:1}) and (\ref{eq:2}) are integrated and averaged along the \textit{y}-direction, during which the velocity profile Eq.~(\ref{eq:6}) is used and the function of \textit{y} in $u'$ is the same as $w'$. This leads to the gap-averaged Navier-Stokes equations
\begin{equation}
\centering
\begin{split}
&\partial_xu+\partial_zw=0,\\
\lambda_j \partial _{t}u=-\frac{1}{\rho}\partial_{x}p-&\frac{2\left ( 1+\mathrm{i} \right )\nu \tanh \left [ \left ( 1+\mathrm{i} \right ) /2\delta_j  \right ] }{b^2\delta_j }u+\frac{1}{\rho}T_x,\\
\lambda_j \partial _{t}w=-\frac{1}{\rho}\partial_{z}p-&\frac{2\left ( 1+\mathrm{i} \right )\nu \tanh \left [ \left ( 1+\mathrm{i} \right ) /2\delta_j  \right ] }{b^2\delta_j }w-g_z\left ( t \right) +\frac{1}{\rho}T_z, \label{eq:33}
\end{split}
\end{equation}
where the nonlinear and viscous terms are neglected but capillary force $(T_x, T_z)$ is included.

If we assume that there is no vortex in the two-dimensional flow,\cite{rajchenbach2011new,li2018faraday,li2019stability,rachik2023effects} a velocity potential $\phi$ is introduced which meets $\partial_x \phi=u$ and $\partial_z \phi=w$. By integrating along the streamline, Eq.~(\ref{eq:33}) is rewritten as
\begin{gather}
\partial_{xx} \phi + \partial_{zz} \phi=0,\label{eq:34}\\
\lambda_j \partial _{t}\phi +\frac{2\left ( 1+\mathrm{i} \right )\nu \tanh \left [ \left ( 1+\mathrm{i} \right ) /2\delta_j  \right ] }{b^2\delta_j }\phi  +g_z\left ( t \right)z -\frac{\sigma }{\rho}\kappa=0, \quad \text{at } z=\zeta.\label{eq:35}
\end{gather}
Using the same method as introduced in Sec.~\ref{sec:level2} to calculate the curvature $\kappa$, Eq.~(\ref{eq:16}) and (\ref{eq:17}) are substituted into Eq.~(\ref{eq:35}), which is then linearized to
\begin{equation}
\lambda_j \partial _{t}\phi +\frac{2\left ( 1+\mathrm{i} \right )\nu \tanh \left [ \left ( 1+\mathrm{i} \right ) /2\delta_j  \right ] }{b^2\delta_j }\phi  +\left ( g-a\cos \omega t \right )\zeta -\frac{\sigma }{\rho}\partial_{xx}\zeta+\frac{2\beta}{\rho b}\partial_t \zeta=0, \quad \text{at } z=0.\label{eq:36}
\end{equation}

To complete the mathematical problem, the impenetrable boundary condition and kinematic boundary condition are included, respectively:
\begin{equation}
\begin{split}
\partial_z \phi=0, \quad &\text{at } z=-\infty,\\
\partial_t \zeta=\lambda_j \partial_z \phi, \quad &\text{at } z=0.\label{eq:37}  
\end{split}
\end{equation}

The solutions for $\phi$ and $\zeta$ have the form of
\begin{equation}
\begin{split}
\phi\left ( x,z,t \right )&=\bar{\phi}\left ( z,t \right )e^{\mathrm{i}kx} ,\\
\zeta \left ( x,t \right ) &=\bar{\zeta} \left ( t \right )e^{\mathrm{i}kx}.\label{eq:38}
\end{split}
\end{equation}
Substituting Eq.~(\ref{eq:38}) into Laplace equation~(\ref{eq:34}), together with the boundary condition~(\ref{eq:37}), the solution of $\bar{\phi}$ is given by
\begin{equation}
\bar{\phi}=\frac{1}{k\lambda_j}\frac{d\bar{\zeta}}{dt}e^{kz}.\label{eq:39}
\end{equation}

Substituting the solutions into Eq.~(\ref{eq:36}), one can obtain the damped Mathieu equation
\begin{equation}
\frac{d^2\bar{\zeta} }{dt^2} +\left \{ \frac{2 \left ( 1+\mathrm{i} \right )\nu\tanh \left [ \left ( 1+\mathrm{i} \right ) /\left ( 2\delta _j \right )  \right ]  }{b^2\delta _j\lambda _j} +\frac{2k\beta }{\rho b}  \right \}\frac{d\bar{\zeta}}{dt} +k\left ( g-a\cos \omega t +\frac{\sigma k^2}{\rho }  \right )\bar{\zeta} =0.\label{eq:40}
\end{equation}
If we set
\begin{equation}
\gamma \ =\gamma_1+\gamma_2,\quad
\gamma_1=\frac{2 \left ( 1+\mathrm{i} \right )\nu\tanh \left [ \left ( 1+\mathrm{i} \right ) /\left ( 2\delta _j \right )  \right ]  }{b^2\delta _j\lambda _j},\quad
\gamma_2=\frac{2k\beta }{\rho b},\quad
\bar{\zeta}=\bar{\eta}e^{-\frac{\gamma t}{2}},
\label{eq:41}
\end{equation}
the standard Mathieu equation is finally derived from Eq.~(\ref{eq:40})
\begin{equation}
\begin{split}
\frac{d^2\bar{\eta}}{dT^2}&+\left (\bar{p}-2\bar{q}\cos 2T\right)\bar{\eta}=0,\\
\bar{p}&=\frac{4k\left (g+\sigma k^2/\rho \right )-\gamma^2}{\omega^2},\\
\bar{q}&=\frac{2ka}{\omega^2},\label{eq:42}
\end{split}
\end{equation}
where $T=\frac{1}{2}\omega t$. The Mathieu equation~(\ref{eq:42}) has been obtained in Refs.~\onlinecite{benjamin1954stability,kumar1994parametric}, and was extended to a generalized Mathieu equation with the inclusion of viscosity by \citet{beyer1995faraday}. 
However, the relevant work is rare in the case of Hele-Shaw cells except the one by \citet{li2018faraday} whereas the capillary force was neglected for simplicity.

To solve the Mathieu equation, we adopt the technique used in Sec.~\ref{sec:level3}. The solution of Eq.~(\ref{eq:42}) is expressed as
\begin{equation}
\bar{\eta}\left ( T \right)=e^{\frac{\gamma}{\omega}T}\sum_{j=1,3,5,\cdots } e^{\mathrm{i} j T}B_j +c.c.,\label{eq:43}
\end{equation}
where $B_j$ have the same meaning as $A_j$ in Eq.~(\ref{eq:24}).

Substituting Eq.~(\ref{eq:43}) into the Mathieu equation~(\ref{eq:42}), a linear equation system for $B_j$ is obtained
\begin{equation}
\begin{split}
&j=1:\quad K_1B_1-aB_1^*-aB_3=0,
\\&j=3:\quad K_3B_3-aB_1-aB_5=0,
\\&j=5:\quad K_5B_5-aB_3-aB_7=0,
\\&j=7:\quad \cdots,
\\ \text{with} \quad &K_j=\frac{\omega^2}{2k}\left (-j^2+\bar{p}+\frac{2\mathrm{i}j\gamma}{\omega}+\frac{\gamma^2}{\omega^2} \right).\label{eq:44}
\end{split}
\end{equation}

Eq.~(\ref{eq:44}) is then recursed to the equation for $j=1$ and solved iteratively to obtain the critical conditions for Faraday instability.

\section{\label{sec:level5}RESULTS AND DISCUSSIONs}

\subsection{\label{sec:level5-1}Comparison with experiments}

To validate the viscous theory developed in the present paper, results are compared with experiments conducted by \citet{li2019stability}. In their experiments, two different Hele-Shaw cells made of Polymethyl Methacrylate (referred to as PMMA) were used, which had a length of 300 mm, a height of 60 mm and a gap size $b=2$ mm or $b=5$ mm. The working liquids were pure ethanol ($> 99.7\ \mathrm{wt} \%$) and a mixture of ethanol and water ($50\  \mathrm{vol} \%$), whose physical properties are given in Table~\ref{tab:1}. 
\begin{table*}
\caption{\label{tab:1}The parameters for the two working liquids. Data are taken from Ref.~\onlinecite{li2019stability}, where the friction coefficient $\beta$ were measured by \citet{hamraoui2000can} }
\begin{ruledtabular}
\begin{tabular}{lcccc}
\mbox{Liquid}&$\mu \ \left ( \mathrm{mPa\ s}  \right ) $&$\rho \left ( \mathrm{g/cm^3}  \right ) $&$\sigma \left ( \mathrm{mN/cm}  \right ) $&$\beta \left ( \mathrm{Pa\ s}  \right ) $\\
\hline
\mbox{Pure ethanol ($> 99.7\ \mathrm{wt} \%$)}&1.096&0.785&0.218&0.040\\
\mbox{Ethanol solution ($50\  \mathrm{vol} \%$)}&2.362&0.926&0.296&0.070\\
\end{tabular}
\end{ruledtabular}
\end{table*}

\begin{figure*}
\subfigure[]{
\includegraphics[height=0.36\linewidth]{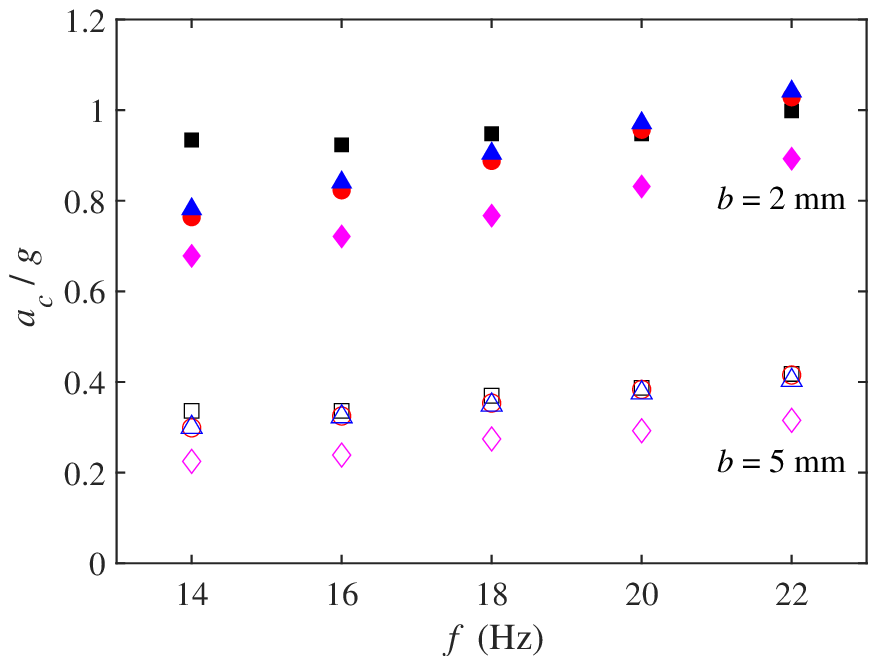}}\quad
\subfigure[]{
\includegraphics[height=0.36\linewidth]{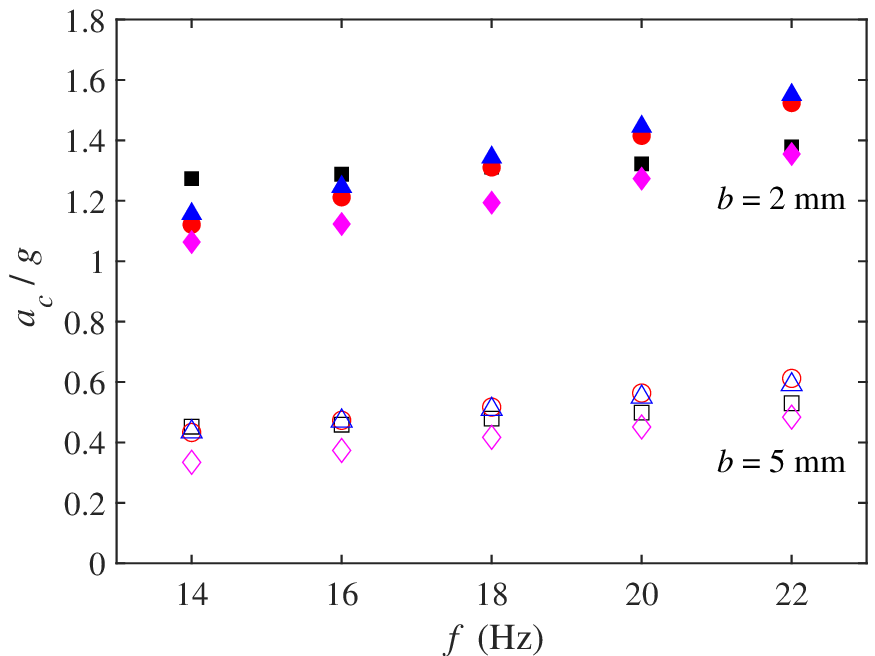}}
\caption{\label{fig:2} Dimensionless critical acceleration amplitude $a_c/g$ for the onset of Faraday waves versus driving frequency \textit{f}. Comparison between theoretical data (circles: results given by present theory; triangles: results given by \citet{bongarzone2023revised}; diamonds: results given by \citet{li2019stability}) and experimental measurements (squares) obtained from \citet{li2019stability}. The filled and empty symbols correspond to $b=2$ mm and $b=5$ mm, respectively. (a) Pure ethanol and (b) ethanol solution.}
\end{figure*}

Let us start with the Faraday instability threshold $a_c$, which predicts the critical acceleration amplitude corresponding to the onset of the free surface. The comparison of dimensionless acceleration amplitude $a_c/g$ against driving frequency \textit{f} is shown in Fig.~\ref{fig:2}, which contains the experimental data along with the results obtained from the present viscous theory and previous theories of \citet{li2019stability} and \citet{bongarzone2023revised} According to Fig.~\ref{fig:2}, theories developed in this paper and by \citet{bongarzone2023revised} give better prediction of critical acceleration than that of \citet{li2019stability}. 
This further illustrates that the approximation of Poiseuille flow underestimates the damping from the gap. 
As has been exhibited by \citet{bongarzone2023revised}, the application of oscillating Stokes boundary layer modifies the damping coefficient $12\nu/b^2$ to a bigger value. 
In this sense, with the influence of periodic parametric excitation being included, the theory of Stokes boundary layer is more suitable for the flow in Hele-Shaw cells. 
In addition, it is obvious that the evaluation is much better for $b=5$ mm than for $b=2$ mm, with the results of ours and \citet{bongarzone2023revised} almost consistent with experimental data for $b=5$ mm. 
\citet{bongarzone2023revised} attributed the disagreement for $b=2$ mm to the phenomenological contact angle model and explained that dissipation from the free surface was more prominent as the gap size \textit{b} decreased. However, with the assumption of a meniscus formed between two lateral walls, theoretical results should have fitted experiments better for $b=2$ mm, since the meniscus is more apparent for smaller gap size. 
In this sense, their explanation is worth deliberating. 
In fact, the free surface undergoes a complicated process that remains unsolved between the lateral walls and existing treatment may be too brief to describe the dynamic behavior of the contact line. 
For instance, according to a numerical simulation conducted by \citet{li2022numerical}, contact-line hysteresis can largely delay the timing of onset. 
A new contact angle model that is more realistic may also benefit the prediction, such as ones proposed by \citet{kidambi2009capillary}, \citet{viola2018capillary}, and so on.
More attention should be paid to this issue to realize a more accurate prediction of Faraday instability in Hele-Shaw cells. 
When it comes to the comparison between the present theory and that of \citet{bongarzone2023revised}, something interesting appears. 
In their derivation, the viscous terms in equations were all neglected after gap-averaging,
but what surprises us is that when viscosity is present the predicted thresholds of Faraday instability are, counterintuitively, slightly smaller than those predicted in the inviscid case.
A detailed explanation of the influence of viscosity will be given in the following subsection, which has not been reported before.\cite{li2019stability,bongarzone2023revised}

\begin{figure*}
\subfigure[]{
\includegraphics[height=0.36\linewidth]{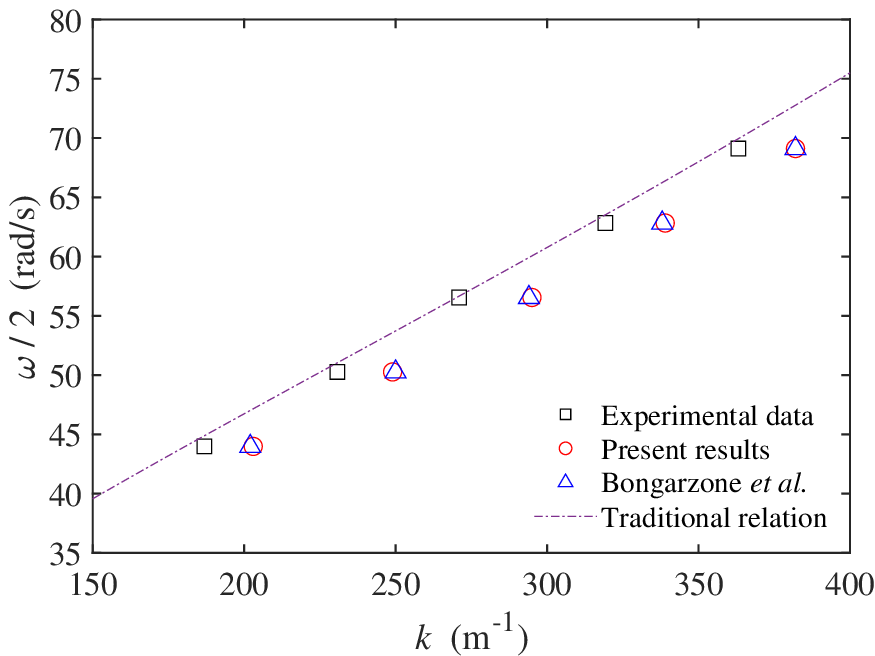}}\quad
\subfigure[]{
\includegraphics[height=0.36\linewidth]{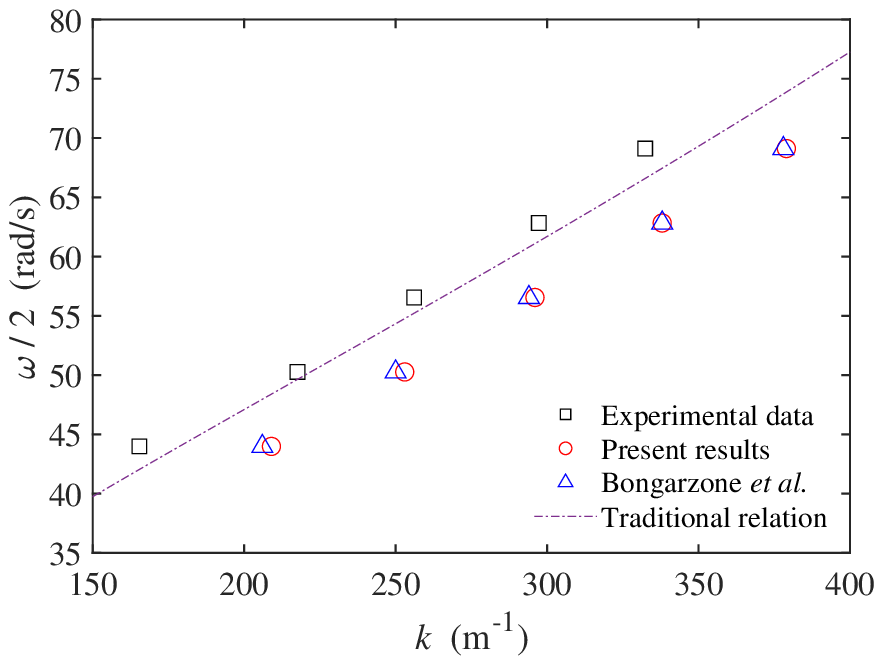}}
\\
\subfigure[]{
\includegraphics[height=0.36\linewidth]{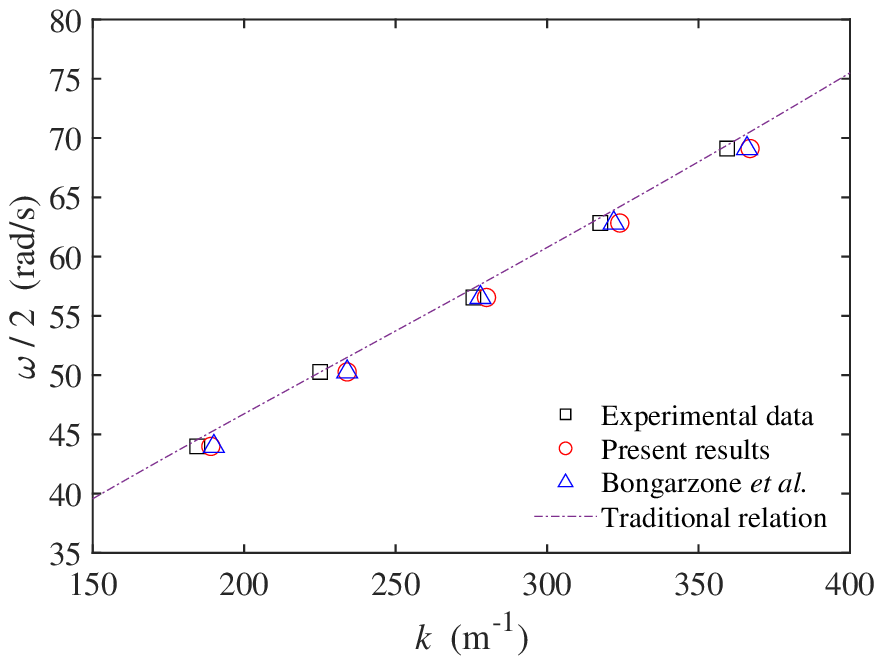}}\quad
\subfigure[]{
\includegraphics[height=0.36\linewidth]{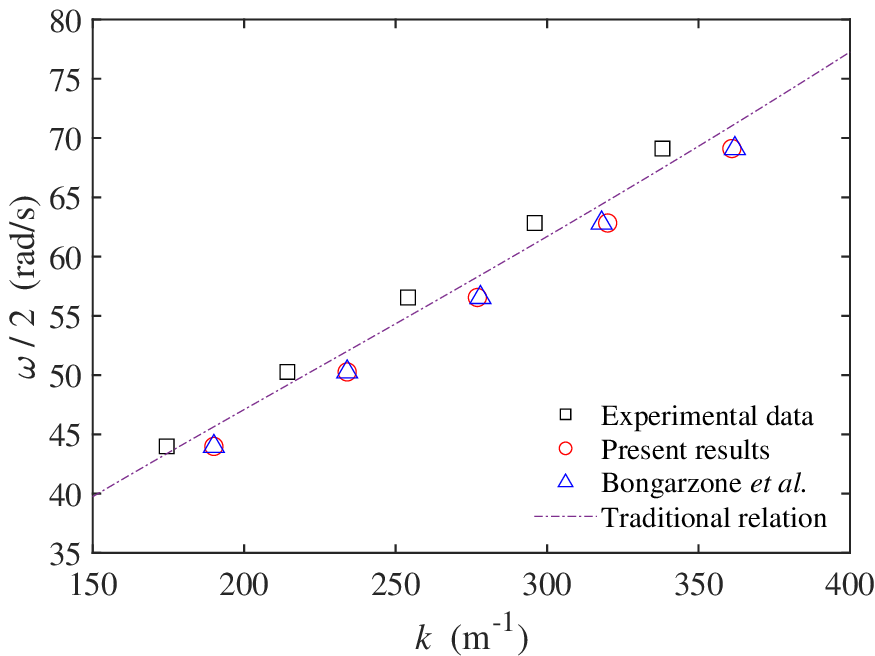}}
\caption{\label{fig:3} Different dispersion relations of Faraday waves. (a) $b=2$ mm for pure ethanol, (b) $b=2$ mm for ethanol solution, (c) $b=5$ mm for pure ethanol, and (d) $b=5$ mm for ethanol solution.}
\end{figure*}

Apart from the critical acceleration amplitude $a_c$, the dispersion relation is another key feature to characterize Faraday instability. 
This is reflected in the critical wavenumber $k_c$ corresponding to the onset. 
Although previous studies have given different modified dispersion relations explicitly or implicitly in the presence of forcing and damping, the traditional linear dispersion relation of Faraday waves
\begin{equation}
\left ( \frac{\omega}{2}\right)^2=gk_0+\frac{\sigma}{\rho}k_0^3\label{eq:45}
\end{equation}
possesses the best agreement to the laboratory observations among others.
Regardless of this, it is not contradictory for us to compare the results obtained from the present theory with experimental data. 
In Fig.~\ref{fig:3}, we plot the experimental data obtained from \citet{li2019stability}, the results given by the present theory and by \citet{bongarzone2023revised}, and traditional dispersion relation Eq.~(\ref{eq:45}) is also included. 
Compared to the traditional dispersion relation, the results of the critical wavenumber given by us and \citet{bongarzone2023revised} both have some deviations, which is called the detuning of the dispersion relation. Since the impact of viscosity on the dispersion relation is relatively small, which can be concluded by comparing our results with those by \citet{bongarzone2023revised}, the detuning may be caused by the gap-averaged damping and the dynamic contact angle. These two factors are what we add to the traditional dispersion relation besides the viscosity. 
This conjecture will be investigated in Subsection~\ref{sec:level5-3}.

\subsection{\label{sec:level5-2}In-plane viscosity}

The straightforward way to look at the impact of viscosity is to compare the full model to the ideal fluid case, while it is prohibited to filter out the viscosity but with rotational motion retained, which has been detailed in the first paragraph in Sec.~\ref{sec:level4}.
Alternatively, to clarify the influence of viscosity on Faraday instability, the potential flow theory has been employed. 
As is known to all, only if fluid is inviscid and flow is irrotational can velocity potential be introduced and potential flow be valid.\cite{lamb1932}
However, if we delve into Eq.~(\ref{eq:27}), there are two terms having something to do with $\nu$.
One is the irrotational part of the viscous dissipation and the other one is the rotational correction.
This distinction between viscous flow and potential flow arises from the combined influence of these two factors.
The following concern is how the two types of physical processes affect the final prediction.

Let us first look at the situation of three-dimensional Faraday instability in the absence of any Darcy effect from Hele-Shaw cells. 
\citet{chen1999amplitude} retained the rotational component during the derivation process. 
A full linear solution was used to obtain the onset condition, and the dimensionless amplitude of driving acceleration $\Delta =ak_0/4\omega_0^2$ was finally expanded as a power series of the damping coefficient associated with viscosity $\tilde{\gamma} =2\nu k_0^2/\omega_0$
\begin{equation}
\Delta_c=\tilde{\gamma}-\frac{1}{2}\tilde{\gamma}^{3/2}+\frac{11-2G}{8\left(3-2G\right)}\tilde{\gamma}^{5/2}+\cdots,\label{eq:46}
\end{equation}
where $k_0$ is the wavenumber obtained from traditional dispersion relation Eq.~(\ref{eq:45}), $\omega_0=\omega/2$ is the wave angular frequency for subharmonics and $G=gk_0/\omega_0^2$ is gravity coefficient (see Ref.~\onlinecite{chen1999amplitude}). 
The first term on the right-hand side is the damping associated with viscosity of fluids, whereas the second term proportional to $\tilde{\gamma}^{3/2}$ is a correction contributed by rotational flow. 
This attribution of correction can be found through the present theory as well. 
Rotational component has an exponent $e^{z\sqrt{k^2+\mathrm{i} j \omega t/2\nu }}$ in the solution given by Eq.~(\ref{eq:27}) and appears with a dimension of $\nu^{3/2}$ in Eq.~(\ref{eq:29}) after expanding, which is similar to the three-dimensional case. 
Now we turn back to Eq.~(\ref{eq:46}), the correction term is of a different order and has a different sign compared to the first term, which means the rotational component reduces the dissipation of the whole system and behaves as an opposite of the irrotational viscid term. 
Similar deductions about the correction of rotational flow were reported in many literature, e.g., the expansion of low viscosity for the linear problem of Faraday waves derived by \citet{muller1997analytic} and the damped Mathieu equation proposed by \citet{nam1993new}.

\begin{figure*}
\subfigure[]{
\includegraphics[height=0.36\linewidth]{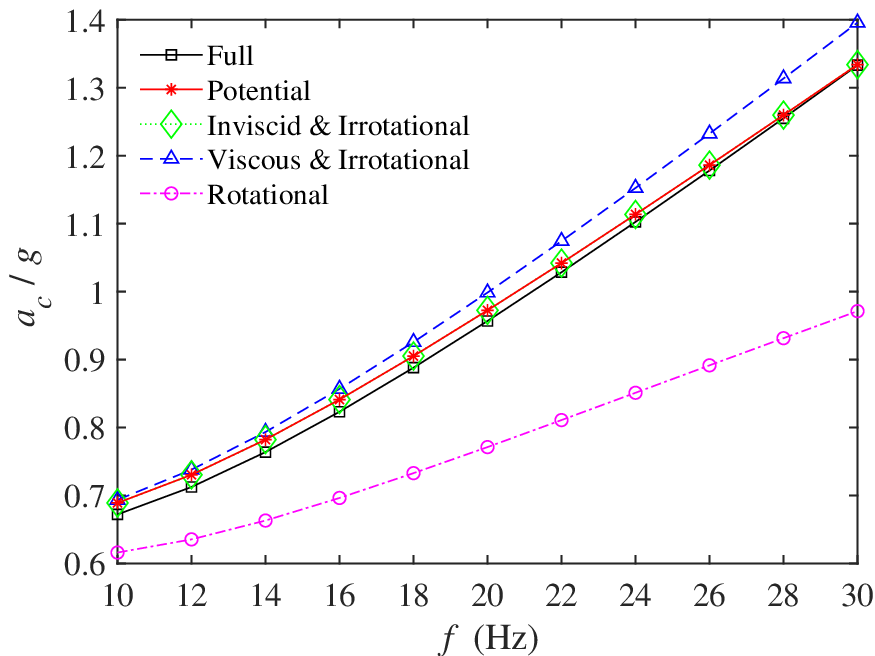}}\quad
\subfigure[]{
\includegraphics[height=0.36\linewidth]{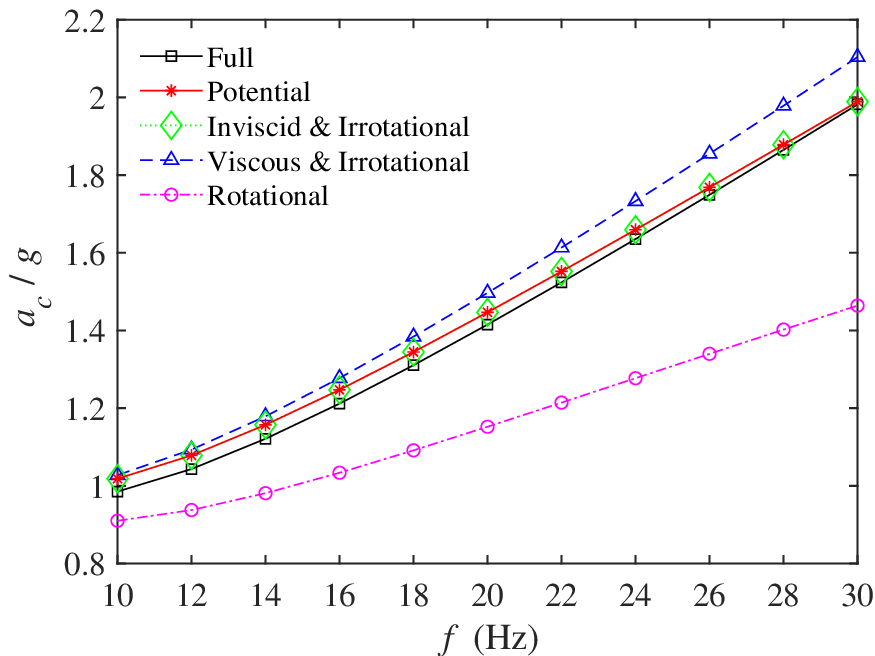}}\\
\subfigure[]{
\includegraphics[height=0.36\linewidth]{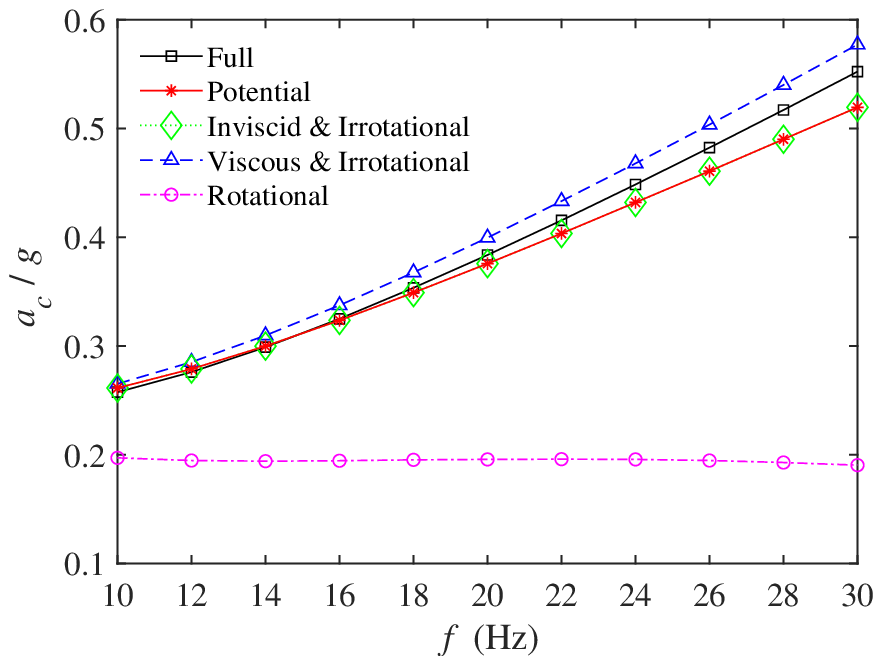}}\quad
\subfigure[]{
\includegraphics[height=0.36\linewidth]{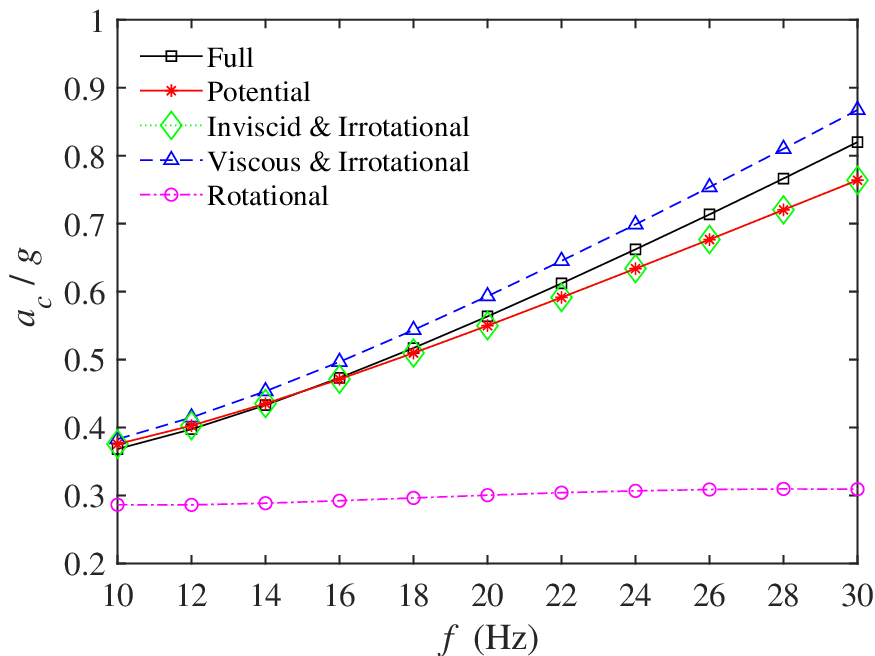}}
\caption{\label{fig:4} Comparison of the theoretical prediction results of dimensionless critical acceleration amplitude. Solid lines with squares: results obtained from the full theory for viscous fluids presented in Sec.~\ref{sec:level2}; solid lines with asterisks: results of potential flow theory in Sec.~\ref{sec:level4}; dotted lines with diamonds: results with the viscous effect being neglected in the full model; dashed lines with triangles: results for viscous \& irrotational case; chain lines with circles: results for rotational case. (a) $b=2$ mm for pure ethanol, (b) $b=2$ mm for ethanol solution, (c) $b=5$ mm for pure ethanol, and (d) $b=5$ mm for ethanol solution.}
\end{figure*}

In Hele-Shaw cells, the terms associated with irrotational viscous dissipation and rotational correction have been separated explicitly in the solution of Eq.~(\ref{eq:27}). 
We hence define the \emph{Viscous \& Irrotational} case and the \emph{Rotational} case respectively as follows. 
For the first case, $w_j(z)$ has the form $\left ( 2k^2\nu +\frac{1}{2} \mathrm{i} j\omega \right ) e^{kz}/\lambda_j$.
The second one is to calculate with $w_j(z)$ of the form $\left(\frac{1}{2}ij\omega e^{kz}-2k^2\nu e^{z\sqrt{k^2 + ijwt/2\nu}}\right)/\lambda_j$ and without consideration of the term $\nu\lambda_j\left(3\partial_{xxz}w +\partial_{zzz}w\right)$ in the normal stress boundary condition Eq.~(\ref{eq:22}) due to its attenuation effect.
Because all theories have already been validated with experiments in the previous subsection, the comparison can be safely extended to a wider range of driving frequencies from 10 to 30 Hz, which contains the main range which is available in experiments.\cite{li2019stability} 
The results of the potential flow theory in Sec.~\ref{sec:level4} and of the full viscous theory are shown in Fig.~\ref{fig:4}. 
As in the previous discussion about the critical acceleration prediction, the same phenomenon emerges that whether considering viscosity or not has a tiny impact on the result. 
\citet{li2019stability} once gave the analysis based on the potential assumption and worried that the viscosity of fluids might be the source of the distinction between theory and experiment. 
From the results given in Fig.~\ref{fig:4}, it can be concluded that obviously when we are only concerned about the threshold of onset, the potential flow theory is feasible, which can simplify the stability analysis to a great extent.
Fig.~\ref{fig:4} also illustrates the results of the \emph{Viscous \& Irrotational} case and the \emph{Rotational} case, along with the inviscid and irrotational one where all terms with $\nu$ involved in Eq.~(\ref{eq:27}) are ignored.
The resulting curves from the last case completely coincide with the one obtained based on potential flow theory, which further validates that the full gap-averaged model can degenerate to the damped Mathieu equation based on the potential flow theory and verifies indirectly the correctness of the derivation in Sec.~\ref{sec:level3}.

From now on the two factors in viscous effect can be discussed with the aid of curves for the \emph{Viscous \& Irrotational} case and the \emph{Rotational} case in Fig.~\ref{fig:4}.
It is obvious that each of them has a remarkable impact on the Faraday instability.
Compared with the inviscid and irrotational case, the thresholds for the appearance of instability plunge largely with only the rotational component involved and are apparently higher in the \emph{Viscous \& Irrotational} case than the reference. 
The amplitude of the decrease is way much greater than the increase, and the overall outcome when combining both factors results in the lower values of the threshold predicted by the full theory when $b=2$ mm compared with the potential flow theory. 
This counterintuitive phenomenon arises here is different from what is observed in the three-dimensional case, where the rotational correction (the second term in the right-hand side of Eq.~(\ref{eq:46})) is smaller than the viscous damping (the first term in the right-hand side of Eq.~(\ref{eq:46})).
Hence, even though the rotational part plays a positive role in triggering instability, the overall effect is stabilizing the flow in a three-dimensional case, but it is another story in Hele-Shaw cells. 
We speculate that the reason may arise from the gap-averaging process, in which the impact of the rotational component can be magnified with the introduction of the gap-averaged damping coefficient. 
One piece of evidence to support this is that for $b=5$ mm as the frequency rises, the curves of the full theory slightly overtop the counterparts of the potential one corresponding to the almost flat lines of the \emph{Rotational} case, which means the gap size is essential in this problem. 
By analogy to the three-dimensional scenario, by expanding Eq.~(\ref{eq:29}), the coefficient of $\nu^{3/2}$ has a different sign compared with the one of $\nu$, but a similar expression like Eq.~(\ref{eq:46}) is hard to derive for the complicated nature of the present model. 
Furthermore, this conclusion is drawn directly from theoretical findings. 
However, validating the rotational influence in-plane in Hele-Shaw cells through experiments is challenging, as separating the rotational component from others requires deliberate design and is currently beyond our capability.

\subsection{\label{sec:level5-3}Damping terms in Mathieu equation}

Within the scope of this study, dissipation in Hele-Shaw system is caused by the in-plane viscosity, the gap-averaged damping, and the dynamic contact angle. 
As discussed previously, the effect of viscosity is relatively small when the rotational component is involved. 
Hence, the main dissipation that dominates the threshold of Faraday instability originates from the gap-averaged damping and the free surface. 
In this subsection, the contribution of these two elements to the critical instability conditions is discussed, where the potential flow theory given in Sec.~\ref{sec:level4} is sufficient.

\begin{figure*}
\subfigure[]{
\includegraphics[height=0.36\linewidth]{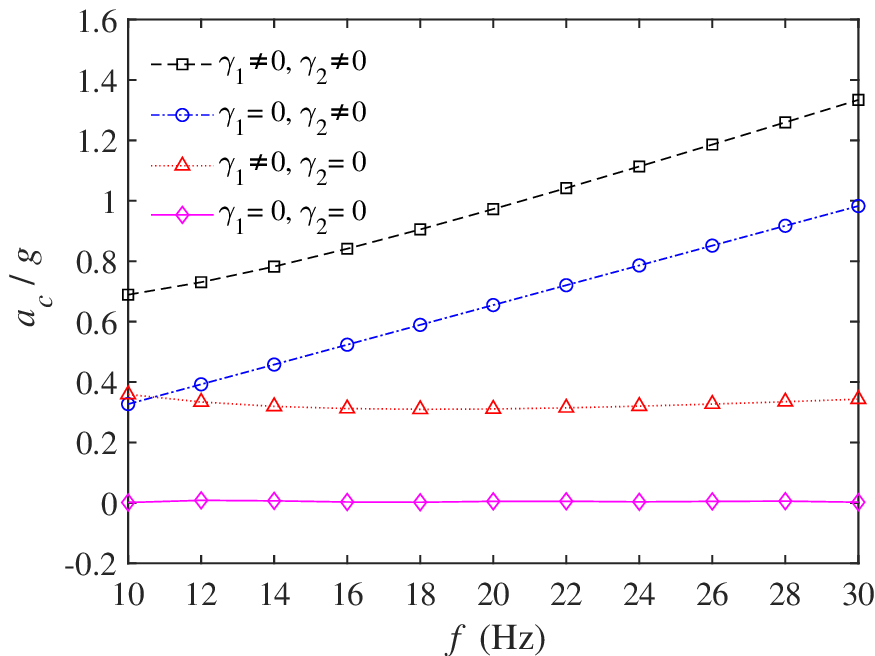}}\quad
\subfigure[]{
\includegraphics[height=0.36\linewidth]{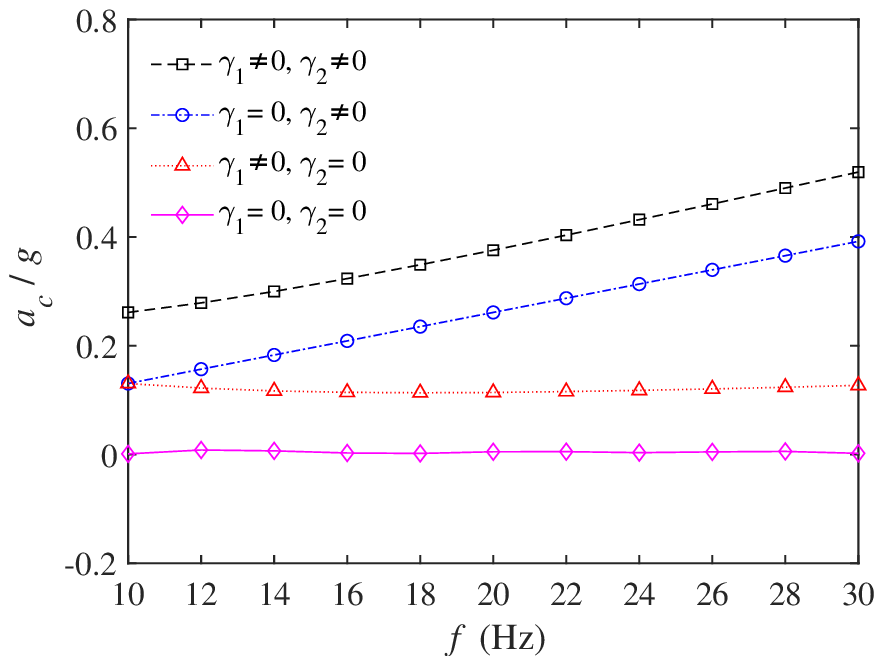}}
\caption{\label{fig:5} The dimensionless critical acceleration amplitudes obtained based on potential approximation versus driving frequency \textit{f}. Comparison is conducted with different damping terms $\gamma_1$ or $\gamma_2$ being incorporated into the damped Mathieu equation Eq.~(\ref{eq:40}). The working fluid is pure ethanol. (a) $b=2$ mm and (b) $b=5$ mm.}
\end{figure*}
The damping terms have been separated individually in Eqs.~(\ref{eq:40}-\ref{eq:41}), where $\gamma_1$ is the damping coefficient derived from the gap-averaged process and $\gamma_2$ is associated with the dynamic contact angle at the free surface. 
Without loss of generality, only results for pure ethanol are discussed here. 
In Fig.~\ref{fig:5}, each combination of the contribution of these two damping terms to the critical acceleration amplitude is plotted. 
Results indicate that the gap-averaged damping $\gamma_1$ provides dissipation that hardly varies with the forcing frequency. 
Although this damping coefficient is associated with the thickness of Stokes boundary layer $\delta=\sqrt{2\nu /\omega}$, where the frequency is included in fact, the main variable that controls the value of this coefficient is the gap size \textit{b} when it is sufficiently small. 
This meets the case of Poiseuille flow, where a damping coefficient $12\nu/b^2$ is obtained. 
When it comes to the dissipation resulting from the dynamic contact angle, Fig.~\ref{fig:5} shows a more important role of this dissipation in the critical acceleration amplitude $a_c$. 
If this factor is neglected, the prediction of $a_c$ is far from the real value. 
Different from the gap-averaged damping, the dissipation from the contact angle increases apparently with forcing frequency \textit{f} rising. 
Although $\gamma_2$ has nothing to do with frequency directly, the wavenumber varies almost linearly with \textit{f} as shown in Fig.~\ref{fig:3}, and $k$ arises in each version of the damping term of the contact angle in Eq.~(\ref{eq:29}) or Eq.~(\ref{eq:40}).
It comes from the out-of-plane curvature estimate with respect to the in-plane velocity.
In reality, it is plausible to picture that the dynamic motion of the contact angle occurs at the same frequency as the wave does and a strong tie between the frequency and the damping term from the contact angle exists.
If we turn to another perspective, it is worth noting that the damping term $\gamma_2$ is sensitive to the value of $\beta$,\cite{bongarzone2023revised} which is obtained from a phenomenological method in experiments, where the physical features of the substrate and the working liquid are essential to measuring this parameter.\cite{hamraoui2000can} 
Due to its sensitivity to $\beta$, A slight discrepancy between the experiments can lead to a significant difference.
This may contribute to the distinction between theory and experiment in the very first subsection in Sec.~\ref{sec:level5}. 
There thus are many issues left worth improving in the dynamic contact angle model.

\begin{figure}
\includegraphics[width=0.7\linewidth]{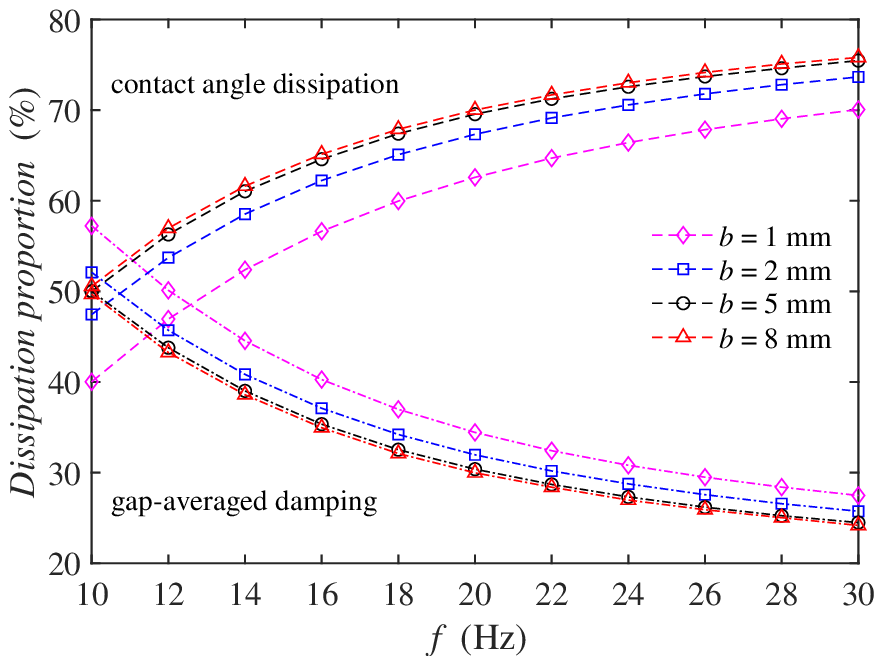}
\caption{\label{fig:6} The proportion of dissipation from dynamic contact angle and gap-averaged damping. Dashed lines (- -): the one associated with dynamic contact angle; chain lines (-- $\cdot$): the one associated with gap-averaged damping.}
\end{figure}
Since the two damping terms in Eq.~(\ref{eq:40}) are added together linearly and the contribution to $a_c$ is independent, we calculate the proportion of each one to the entire dissipation. 
We extend the range of gap size to reveal the relationship between dissipation and geometric features of Hele-Shaw cells.  
A tendency appears similar to Fig.~\ref{fig:5} that dissipation caused by dynamic contact angle is strengthened as the forcing frequency increases. 
However, $\gamma_1$ experiences a decline as $f$ escalates, which differs from what is observed in Fig.~\ref{fig:5}.
We argue that $\gamma_2$ stabilizes the liquid motion in a more direct way than $\gamma_1$ does, given the different performances between Fig.~\ref{fig:5} and Fig.~\ref{fig:6} as $f$ changes.
As for the influence of the gap size on the distribution of the two ingredients,
according to Fig.~\ref{fig:6}, with the gap size decreasing, the influence of $\gamma_1$ becomes more prominent compared to $\gamma_2$, which means the dissipation from two lateral walls is enlarged, despite the lower level from $\gamma_1$ than its counterpart most of the time.
There seems to be a convergent percentage for each of them when the gap size is large enough if we look at cases with $b=5$ mm and $b=8$ mm closely. 
This is a strong conclusion.
Nevertheless, as the container tends to be a wider one the three-dimensional nature appears,
and then the assumption on which our theory is based is gradually ruined.
Dissipation from all walls matters in that scenario, which will leave this convergence meaningless.
To fully exploit this trend of convergence one needs to find the upper bound of the gap size as a necessity.

\begin{figure*}
\subfigure[]{
\includegraphics[height=0.36\linewidth]{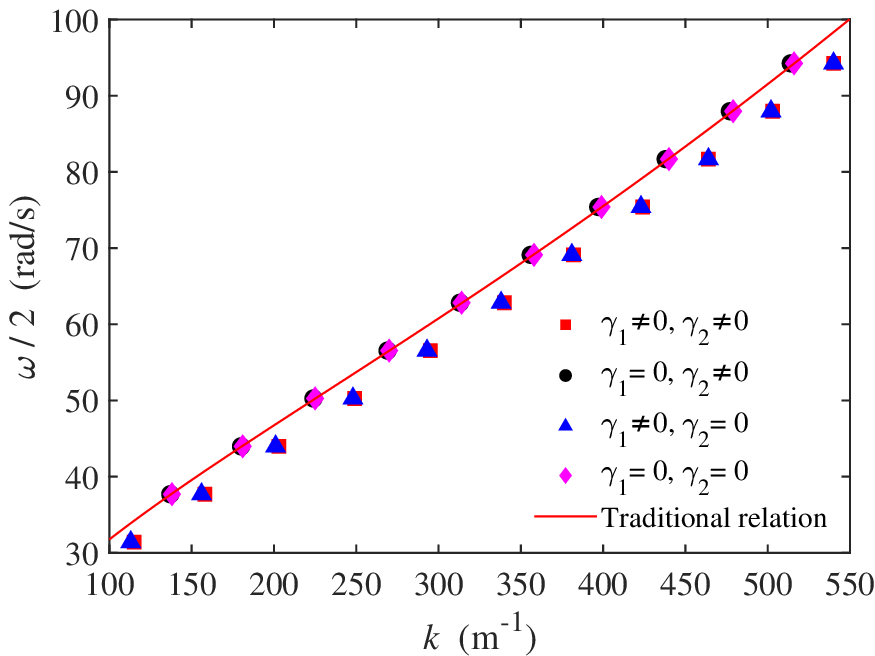}}\quad
\subfigure[]{
\includegraphics[height=0.36\linewidth]{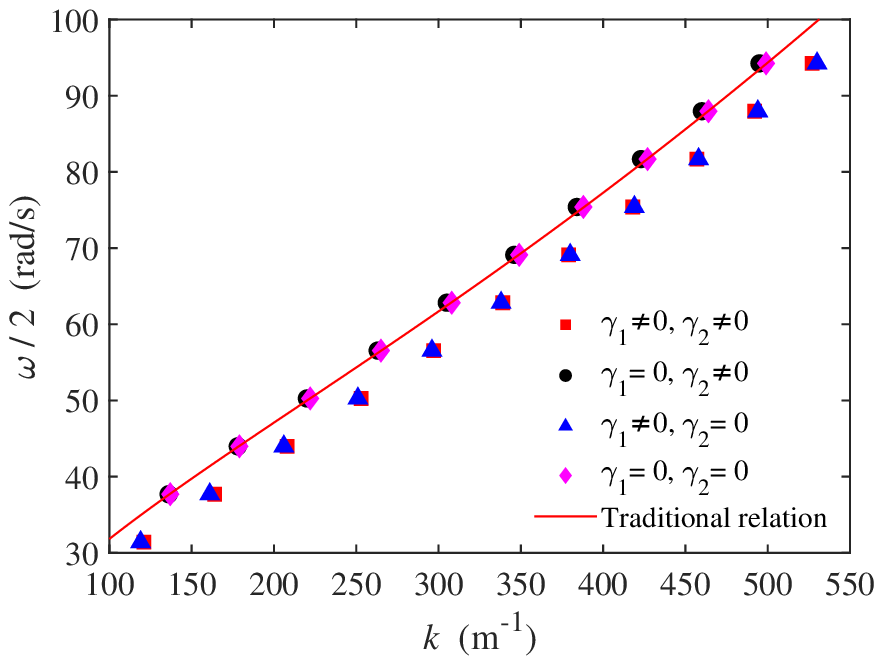}}
\caption{\label{fig:7} The dispersion relation of Faraday waves in Hele-Shaw cells. Comparison between traditional dispersion relation and potential theory with the damped terms $\gamma_1$ or $\gamma_2$ in Eqs.~(\ref{eq:40}-\ref{eq:41}) being neglected. The gap size is $b=2$ mm. (a) Pure ethanol and (b) ethanol solution.}
\end{figure*}
In the previous subsection, we speculate that $\gamma_1$ and $\gamma_2$ should be blamed for the detuning of the dispersion relation, which is manifested in the error bar shown in Fig.~\ref{fig:3}. 
To analyze this problem, 
comparison between results obtained from each combination of the two terms and traditional dispersion relation is made in Fig.~\ref{fig:7}. 
Since detuning is prominent in cases of $b=2$ mm, without loss of generality, only results for this gap size are shown. 
According to Fig.~\ref{fig:7}, the two cases with the gap-averaged damping coefficient $\gamma_1$ being removed coincide with the traditional dispersion relation quite well, while the cases with this term being considered show a deviation. 
In contrast, the influence of dynamic contact angle on the dispersion relation is negligible. 
In this sense, the detuning comes mainly from the gap-averaged damping coefficient. 
This detuning has been observed by \citet{bongarzone2023revised}, whose results are similar to ours as shown in Fig.~\ref{fig:3}. 
According to their statement, the real part of the damping coefficient $\gamma_1$ leads to a higher value of critical acceleration amplitude, while the imaginary part causes the detuning. 
This reason is confirmed in the present work by removing the imaginary part of the gap-averaged damping coefficient, and the prediction is much better just as the ones shown in Fig.~\ref{fig:7} when $\gamma_1$ is neglected. 
We know that so far although the application of Stokes boundary layer enhances the prediction of instability threshold, the unreal detuning shows up. 
It brings in an inherent contradiction for this updated out-of-plane velocity profile.
If the improvement of the estimate of the damping comes at the cost of hindering the dispersion relation, the gap-averaged damping must be overestimated as well.
Considering the negligible impact of viscous dissipation discussed earlier, the contact angle model remains the sole viable option to mitigate the instability. 
However, the current formulation of $\gamma_2$ fails to adequately offset the potentially overestimated value of $\gamma_1$.
In spite of the promising results of the present study, a more accurate model is still needed to precisely solve this problem in the future.

\section{\label{sec:level6}CONCLUSIONS}

In the scenario of three-dimensional Faraday waves, viscous theory has been widely used to predict the instability threshold and pattern formation beyond onset.\cite{kumar1994parametric,chen1999amplitude} 
But in Hele-Shaw cells, existing theories are all based on potential flow theory or neglect the viscosity of fluids after gap-averaging. 
The effect of viscosity on this problem remains unknown, we thus apply viscous theory to the study of Faraday instability in Hele-Shaw cells.

In this paper, a new model that contains the viscosity of fluids, the surface tension, and the dynamic contact angle is derived to predict the critical conditions for the onset of Faraday waves in Hele-Shaw cells. 
Starting from three-dimensional Navier-Stokes equations, the governing equation is simplified with viscosity being reserved. 
An ingenious gap-averaging technique is significant since viscosity complicates the problem largely. 
The gap-averaged damping and the effect of viscosity are clarified during the derivation process. 
After expanding the equations and removing nonlinear terms, the final linear hydrodynamic model is obtained to describe the Faraday instability. 
To access the critical acceleration amplitude and critical wavenumber, a semi-analytical method\cite{chen1999amplitude} is used and a system of equations for the amplitude of the solution is solved recursively. 
Finally, results of the present theory are compared with experiments conducted by \citet{li2019stability} and with some previous theoretical ones.

We conclude that the presence of viscosity leads to a small correction compared to the potential flow theory. 
Although the viscosity has no apparent impact on the onset as a whole, the irrotational and viscous component and the rotational correction have different effects on the critical conditions for the onset. 
The rotational correction can trigger the motion rather than suppress it to a large extent. 
Each of them is qualitatively consistent with the three-dimensional case but the overall consequence is counterintuitive compared to the three-dimensional scenario. 
If the detailed influence of viscosity is not the priority, the potential flow theory is sufficient to predict the onset in Hele-Shaw cells. 
In this sense, a novel damped Mathieu equation is obtained with two damping terms referring to the gap-averaged damping and energy loss from the dynamic contact angle. 
The gap-averaged damping is enhanced as the gap size decreases while dissipation from the dynamic contact angle grows for higher frequency. 

Despite the promising outcome given in the present study, no theory can provide an exact agreement to the experimental observation of the onset, especially when the gap size turns smaller.
As for the dispersion relation, the unusual detuning appearing in the present theory comes from the updated damping coefficient. 
The introduction of the theory of Stokes boundary layer leads to a contradiction between a greater dissipation from two lateral walls and a deviation from the real dispersion relation, which means it probably overestimates the damping as well.
If this is the case, the offset introduced by this misuse should be compensated by the dissipation of the out-of-plane capillary force.
Actually, there is room to improve the theory since the calculation of the curvature based on molecular dynamics is simply a linear model and relies on the phenomenological parameter heavily.
In addition, the existence of the liquid film on lateral walls makes the wetting process more complicated which has not been considered by any existing literature yet.\cite{li2019stability} 
Considering the lack of experimental records revealing the real dynamic behavior of the out-of-plane free surface, the treatment of the contact line deserves more attention to modify the current models.

\begin{acknowledgments}
The authors are grateful to the private communication with Professor Jorge Vi\~{n}als. The authors would like to acknowledge the National Natural Science Foundation of China (Approval No.12002206) and Fundamental Research Funds for the Central Universities for the financial support of this work.
\end{acknowledgments}

\section*{Declaration of Interests}
The authors report no conflict of interest.

\section*{Data Availability}
The data that supports the findings of this study are available within the article and from the corresponding author upon reasonable request.

\appendix

\section{\label{sec:levelA1}The oscillating Stokes boundary layer}
The oscillating Stokes boundary layer introduced in this paper is inspired by \citet{bongarzone2023revised} Here we give a rough description of Eq.~(\ref{eq:6}). For the original derivation please refer to \citet{bongarzone2023revised}

Linearizing Eq.~(\ref{eq:1}) at the rest state $\bm{u}'=0$ and $p=-\rho g_z\left ( t \right )z$, one can obtain
\begin{equation}
\frac{\partial \bm{u}'}{\partial t} =-\frac{1}{\rho } \nabla p+\nu \nabla^2\bm{u}'.
\label{appeq:1}
\end{equation}
Assuming that $bk\ll 1$, the velocity through the gap satisfies $v'\ll u',w'$. Then Eq.~(\ref{appeq:1}) can be simplified as
\begin{equation}
\frac{\partial u'}{\partial t} =-\frac{1}{\rho} \frac{\partial p}{\partial x}+\nu \frac{\partial^2 u'}{\partial y^2},\quad
\frac{\partial w'}{\partial t} =-\frac{1}{\rho} \frac{\partial p}{\partial z}+\nu \frac{\partial^2 w'}{\partial y^2},\quad
\frac{\partial p}{\partial y}=0.
\label{appeq:2}
\end{equation}
By making dimensionless as follows:
\begin{equation}
\bar{x}=xk, \quad
\bar{y}=\frac{y}{b}, \quad
\bar{z}=zk, \quad
\bar{u}=\frac{u'}{a\omega}, \quad
\bar{v}=\frac{v'}{bka\omega}, \quad
\bar{w}=\frac{w'}{a\omega}, \quad
\bar{p}=\frac{kp}{\rho a \omega^2}, \quad
\bar{t}=\omega t,
\label{appeq:3}
\end{equation}
the first two equations in Eq.~(\ref{appeq:2}) can be expressed in a dimensionless form which is written as
\begin{equation}
\frac{\partial \bar{u}}{\partial \bar{t}} =-\frac{\partial \bar{p}}{\partial \bar{x}}+\frac{\delta_{St}^2}{2} \frac{\partial^2 \bar{u}}{\partial \bar{y}^2}, \quad
\frac{\partial \bar{w}}{\partial \bar{t}} =-\frac{\partial \bar{p}}{\partial \bar{z}}+\frac{\delta_{St}^2}{2} \frac{\partial^2 \bar{w}}{\partial \bar{y}^2},
\label{appeq:4}
\end{equation}
where $\delta_{St}=\sqrt{2\nu /\omega } /b$ is the dimensionless thickness of the oscillating Stokes boundary layer. 

When separating the terms relevant to $\bar{t}$, the solutions of Eq.~(\ref{appeq:4}) are written as $\bar{\bm{u}}\left ( \bar{\bm{x}}; \bar{t} \right )=Re\left [\tilde{\bm{u}} \left ( \bar{\bm{x}}\right ) e^{\mathrm{i}  j \bar{t}/2} \right ]$, where $j=1,3,5,\cdots$ denote Fourier terms. Substituting it into Eq.~(\ref{appeq:4}), it can be found that each Fourier component must satisfy
\begin{equation}
\frac{j \mathrm{i}}{2}\tilde{u} =-\frac{\partial \bar{p}}{\partial \bar{x}}+\frac{\delta_{St}^2}{2} \frac{\partial^2 \tilde{u}}{\partial \bar{y}^2},\quad
\frac{j \mathrm{i}}{2}\tilde{w} =-\frac{\partial \bar{p}}{\partial \bar{z}}+\frac{\delta_{St}^2}{2} \frac{\partial^2 \tilde{w}}{\partial \bar{y}^2}.
\label{appeq:5}
\end{equation}
By solving the ordinary differential equations about $\bar{y}$, along with the no-slip condition at $\bar{y}=\pm \frac{1}{2}$, the explicit functions of $\bar{y}$ are separated that
\begin{equation}
\tilde{u}=\frac{2\mathrm{i} }{j}\frac{\partial \bar{p}}{\partial \bar{x}}\left \{ 1-\frac{\cosh \left [ \left ( 1+ \mathrm{i} \right ) \bar{y}/\delta _j \right ] }{\cosh \left [ \left ( 1+\mathrm{i} \right )/2\delta _j \right ]}  \right \}, \quad
\tilde{v} =0, \quad
\tilde{w}=\frac{2\mathrm{i} }{j}\frac{\partial \bar{p}}{\partial \bar{z}}\left \{ 1-\frac{\cosh \left [ \left ( 1+ \mathrm{i} \right ) \bar{y}/\delta _j \right ] }{\cosh \left [ \left ( 1+\mathrm{i} \right )/2\delta _j \right ]}  \right \}.
\label{appeq:6}
\end{equation}
The coefficient $\delta_j$ is somewhat different from the one in Ref.~\onlinecite{bongarzone2023revised}, since different forms of solutions are assumed. Using the function of $\bar{y}$ in Eq.~(\ref{appeq:6}) to describe the velocity profile leads to the following expression in a dimensional form:
\begin{equation}
\begin{split}
u'\left ( x,y,z,t \right ) &=\left \{ 1-\frac{\cosh \left [ \left ( 1+\mathrm{i} \right ) y/(b\delta _j) \right ] }{\cosh \left [ \left ( 1+\mathrm{i} \right )/2\delta _j \right ] }  \right \} u\left ( x,z,t \right ),\\
v'\left ( x,y,z,t \right ) &=0,\\
w'\left ( x,y,z,t \right ) &=\left \{ 1-\frac{\cosh \left [ \left ( 1+\mathrm{i} \right ) y/(b\delta _j) \right ] }{\cosh \left [ \left ( 1+\mathrm{i} \right )/2\delta _j \right ] }  \right \} w\left ( x,z,t \right ).
\end{split}
\label{appeq:7}
\end{equation}

\nocite{*}
\bibliography{aipsamp}

\end{document}